# The Linear-Time-Invariance Notion of the Koopman Analysis—Part 1: The Architecture, Practical Rendering on the Prism Wake, and Fluid-Structure Association

Cruz Y. Li[12] (李雨桐), Zengshun Chen[1*] (陈增顺), Tim K.T. Tse[3**] (谢锦添), Asiri Umenga Weerasuriya[4], Xuelin Zhang[5] (张雪琳), Yunfei Fu[6] (付云飞), Xisheng Lin[7] (蔺习升)

*[1] Department of Civil Engineering, Chongqing University, Chongqing, China*

*[2,3,4,6,7] Department of Civil and Environmental Engineering, The Hong Kong University of Science and Technology, Hong Kong SAR, China*

*[5] School of Atmospheric Sciences, Sun Yat-sen University, Zhuhai, China.*

[1] yliht@connect.ust.hk; ORCID 0000-0002-9527-4674

[2] zchenba@connect.ust.hk; ORCID 0000-0001-5916-1165

[3] timkttse@ust.hk; ORCID 0000-0002-9678-1037

[4] asiriuw@connect.ust.hk; ORCID 0000-0001-8543-5449

[5] zhangxlin25@mail.sysu.edu.cn; ORCID 0000-0003-3941-4596

[6] yfuar@connect.ust.hk; ORCID 0000-0003-4225-081X

[7] xlinbl@connect.ust.hk; ORCID 0000-0002-1644-8796

[*] Co-first author with equal contribution.

[**] Corresponding author

All correspondence is directed to Dr. Tim K.T. Tse.

# Abstract

This work proposes a Linear-Time-Invariance (LTI) notion to the Koopman analysis, finding an invariant subspace on which Koopman modes are consistent and physically meaningful. It also develops the Koopman-LTI architecture---a systematic procedure to associate fluid excitation and structure surface pressure by matching Koopman eigen tuples, solving a longstanding problem for fluid-structure interactions. The architecture is data-driven and modular, accommodating all types of data and Koopman algorithms. Through a pedagogical demonstration on a prism wake and the rudimentary Dynamic Mode Decomposition algorithm, results show a near-exact linearization of nonlinear turbulence, with mean and rms errors of $O^{-12}$ and $O^{-9}$, respectively. The DMD also approximated the Koopman modes with $O^{-8}$ error. The LTI reduced the subcritical prism wake during shear layer transition II into only six dominant excitation-response Koopman modal duplets. The upstream and crosswind walls constitute a dynamically unified interface dominated by only two mechanisms. The downstream wall remains a distinct interface and is dominated by four other mechanisms. The complete revelation of the prism wake essentially comes down to understanding the six mechanisms, which Part 2 (*Li et al.*, 2022) will address by investigating the physical interpretations of the duplets' in-synch, phenomenological features. Finally, the current analysis also revealed *w*'s trivial role in this convection-dominated free-shear flow, Reynolds stresses' spectral description of cascading eddies, vortices' sensitivity to dilation and indifference to distortion, and structure responses' origin in vortex activities.

# 1. Introduction

Recent advances in data science and computational hardware brought about practical resuscitations to several long-established mathematical wonders. The Koopman theory is a major one in fluid mechanics (Koopman, 1931; Koopman & Neumann, 1932). In essence, it proposes to compensate nonlinear losses by dimensional gains. Seven decades later, Mezić (2005) brought life to the theory by developing a data-driven approximation of the infinite-dimensional Koopman operator on finite manifolds. Popular Koopman algorithms include the Generalized Laplace Analysis (GLA) (Mauroy & Mezić, 2012, 2013, 2016), the Ulam Galerkin method (Bollt & Santitissadeekorn, 2013; Froyland et al., 2014), the Dynamic Mode Decomposition (DMD) (Rowley et al., 2009; Schmid, 2010), the Spectral Proper Orthogonal Decomposition (SPOD) (Schmidt & Colonius, 2020; Towne et al., 2018), and some machine-learning networks (Brunton et al., 2020; Lusch et al., 2018; M Raissi et al., 2019; Rudy et al., 2017). Budišić et al. (2012) termed this collective effort as applied Koopmanism.

Like any other data-driven technology, applied Koopmanism has algorithmic and analytical ends. The former works toward the idealization of methods, and the latter focuses on the interpretation of algorithmic outcomes. To date, most research is dedicated to algorithmic development, and the interpretation side is often left to individual efforts with less collectiveness and depth. The main reason is that Koopman/Fourier eigen tuples are input-dependent, constantly changing with uncertain implications. The *a posteriori* evidence in this work suggests that upon finding an invariant Koopman subspace (pragmatic), Koopman mode shapes contain similar coherent structures, therefore, consistent, useful information.

Specifically, the investigation was conducted through the scope of fluid-structure interaction (FSI). Until now, even with deterministic equations for macroscopic fluid motions, the Navier-Stokes equations, FSI is still an unsolved problem, especially with turbulence (Pope, 2000;

Tennekes & Lumley, 2018). We usually find associating fluid excitations with structural responses immensely difficult. Although many have applied the Koopman analysis, or its algorithmic subordinates, the Dynamic Mode Decomposition (DMD), to FSI systems (Basley et al., 2013; Dotto et al., 2021; Eivazi et al., 2020; Garicano-Mena et al., 2019; Gómez et al., 2014; Higham et al., 2021; Jang et al., 2021; Kemp et al., 2021; N.-H. Liu et al., 2021; Y. Liu, Huang, et al., 2021; Y. Liu, Long, et al., 2021; Ping et al., 2021; Sun et al., 2021; Yuan et al., 2021), none tried establishing a fluid-structure relationship by matching Koopman tuples. It is now possible with consistent Koopman modes, and the relationship has tremendous value for understanding FSI. It supports the interpolation, extrapolation, and prediction of fluid dynamics regardless of technical finesse (He et al., 2022; Lusch et al., 2018; Maziar Raissi et al., 2019). It also bears implications in essentially every scientific discipline: acoustics and vibrations, hazards resilience, aeronautics, wind-hydro energy, cardiology, and so on.

Summarizing experimental results, this serial effort proposes a Linear-Time-Invariance (LTI) notion to the original Koopman analysis, or the Koopman Linear-Time-Invariant (Koopman-LTI) architecture. Its significance is summarized as follows:

1. A new perception employing Koopman eigen tuples to associate fluid and structure for physical insights.
2. A systematic procedure to acquire sampling-independent Koopman modes, relate the flow field excitation and structure surface pressure, and uncover underlying FSI physics.
3. A spectral characterization and phenomenological study of the primary test subject, the prism wake, which bears broad pedagogical and engineering implications.

The compositional sequence of this paper can be summarized as follows. Section 2 describes the Koopman-LTI architecture. Section 3 offers a practical rendering of the architecture and establishes the LTI notion. Section 4 spectrally characterizes the prism wake and underpins the prominent fluid-structure relationships. Section 5 presents the summary. The follow-up work, as Part 2, studies the Koopman modes' physical interpretations, analyzes the phenomenology of the fluid-structure duplets, and reveals the major FSI mechanisms.

We also point out a limitation of this work. Its conclusions are based on empirical evidence and direct observations, and the mathematical underpinnings are beyond the authors' expertise. On this note, we invite expert opinions and future efforts to grace the brute force results with mathematical elegance.

# 2. The Koopman Linear-Time-Invariant Architecture

## *2.1 The Koopman Operator Theory*

In early 1930s, B. O. Koopman (1931; 1932) outlined the possibility of representing a nonlinear dynamical system in terms of an infinite-dimensional Hamiltonian, which is a linear operator that acts on the Hilbert space of measurement functions of the system's state (Brunton, 2019).

Following Mezić (2005) and Rowley *et al.* (2009), one may consider a dynamical system in discrete time,

$$\boldsymbol{y}_{i+1} = \boldsymbol{f}(\boldsymbol{y}_i), \tag{1}$$

where $i \in \mathbb{Z}$ and $\boldsymbol{f}$ is a map from a manifold $M$ to itself. The Koopman operator $U$ is linear, infinite-dimensional, and acts on scalar-valued functions on $M$. For any scalar-valued function $g: M \to \mathbb{R}$, $U$ maps $g$ into a new function (Rowley et al., 2009),

$$U\, g(\boldsymbol{y}) = g(\boldsymbol{f}(\boldsymbol{y})). \tag{2}$$

For a linear system, $U$ is exact. For a nonlinear system, it is a linearization of nonlinearities. Conceptually, one may think of $U$ as a discretization of a curve: as the total number of linear segments approaches infinity, the linear approximation approaches continuity and perfection. Therefore, $U$ is the globally optimal linearization.

### *2.2 Data-Driven Koopmanism*

For its dimensionality, acquiring the full-order $U$ is only a theoretical possibility. The works of Mezić and colleagues bear milestone significance precisely because they formulated a data-driven approach to approximate the infinite operator on finite-dimensional subspaces---the practical realization of the promising theory (Budišić et al., 2012; Froyland et al., 2014; Mauroy & Mezić, 2013; Rowley et al., 2009; Sayadi et al., 2014).

As mentioned above, there are many algorithms for computing the finite-dimensional approximation and the Koopman eigen tuples (*i.e.*, eigenfunctions, eigenvalues, Koopman modes). The GLA, with the prescription of eigenvalues, approximates the Koopman modes and eigenfunctions (Budišić et al., 2012; Mauroy & Mezić, 2013). The Ulam Galerkin method does so for the eigenfunctions and eigenvalues in the approximation of the Perron-Frobenius operator, which is the adjoint Koopman operator (Froyland et al., 2014). The DMD approximates the eigenvalues and Koopman modes with algorithmic simplicity and robustness (Tu et al., 2014; Williams et al., 2015). The SPOD, in mathematical essence, is the ensemble average of the DMD (Towne et al., 2018), so it inherits the Koopman implications. The list also includes all variants of the vanilla algorithms, and even some machine (Q. Li et al., 2017; Pan et al., 2021) and machine learning (Brunton et al., 2020; Lusch et al., 2018) techniques.

### *2.3 The Koopman Linear-Time-Invariance Architecture*

Following its prelude (C. Y. Li et al., 2021), the present work constructs the complete, modular architecture of the Koopman-LTI (see figure 1). This architecture founds upon three core principles:

1. On an invariant Koopman subspace, where Koopman eigen tuples are independent of input samples, a mode's bin-wise averaged coherent shape and a sufficiently resolved discrete spectrum contain meaningful physical implications.
2. Linear quantities can be added, subtracted, and compared directly on the same Koopman subspace (discrete spectrum).
3. The fundamental awareness that no matter how complex, unsolvable, or even undiscernible the governing equations may be, the fluid and the structure must somehow correlate to each other and conform to some consistent laws and underlying physics. This underlying message is embedded in the data and measurements. A sampling-independent LTI model is the globally optimal linearization of this information.

On this note, the Koopman-LTI is not a new decomposition algorithm---we give full credit to the giants who have developed the theories and algorithms, and on whose shoulders we stand (Brunton et al., 2020; Brunton, Proctor, et al., 2016; Budišić et al., 2012; Koopman, 1931; Kutz et al., 2016; Mauroy & Mezić, 2013; Mezić, 2005; Rowley et al., 2009; Schmid, 2010). It is a mode of thinking, an analytical procedure set to manipulate, articulate, and interpret algorithmic outcomes. Its significance reflects on in-depth analysis of Koopman eigen tuples, which, for our specific agenda herein, enables a deterministic fluid-structure association and enhanced understandings of FSI.

On a secondary level, the Koopman-LTI also introduces several methodical improvements:

1. A set of evaluation metrics and practice guidelines to achieve sampling independence for the DMD, guaranteeing a temporally converged LTI model that captures all the long-term, recurring dynamics of the input data.

2. The dynamic Koopman mode shape, which not only includes phase information but also facilitates the in-synch comparison of the flow field excitation and structure response.

The Koopman-LTI consists of five modules: *Input Curation*, *Koopman Algorithm*, *Linearly-Time-Invariance*, *Constitutive Relationship*, and *Phenomenological Relationship*.

### *2.3.1 Input Curation*

The *Input Curation* module sorts and pre-processes the input data. The input data, whether by field, experimental, or numerical techniques, can take in any representative variables as independent realizations. No prior processing is required to enhance the physical or algorithmic connections. Of course, the input data's quality decides the modeling accuracy, which shall be best warranted. Noise treatment and interpolation may also be necessary for highly contaminated raw measurements.

### *2.3.2 Koopman Algorithm*

The *Koopman Algorithm* module is the algorithmic staple and received the most attention. This stage performs the global linearization and decomposes the input sequences into linear superpositions of Koopman eigen tuples. The decomposition does not require the solution or, more impressively, any knowledge of the input system. And, theoretically, any algorithm approximating the Koopman eigen tuples works for this module (Budišić et al., 2012; Mezić, 2005; Williams et al., 2015). The input data format shall adhere to the algorithm's requirement.

### *2.3.3 Linear-Time-Invariance*

The following modules are beyond algorithm development. The *Linear-Time-Invariance* module centers on the LTI notion, which is reflected by an invariant Koopman subspace, or a converged Koopman model that captures all the long-term, recurring dynamics. One shall reckon the LTI does not abide the theoretical dogma outlined in Brunton et al. (2016). Instead, it adheres to the pragmatic, empirical observations and guidelines in Li et al. (2022a).

Therein, the *Linearity* part refers to the Koopman linearization, its accuracy, stability, and causality. The *Time-Invariance* part refers to sampling independence. If the Koopman linearization is sensitive to input changes (pragmatic because true independence only exists in theory when discrete bins become continuous), the model captures only a part of the full state space. Whether captured portion is non-trivial to the overall slow subspace is also completely unwarranted (Williams et al., 2015). Therefore, sampling independence is an indispensable pillar to the ensuing fluid-structure association.

A fitting analogy for the Koopman analysis is a photograph. It maps the nonlinear features onto a linear space as a combination of linear elements, the pixels. Photos are sensitive to angles (input-dependence). The LTI notion turns the photo into a 3D scan. With it, one captures all the steady dynamics regardless of the object's translation and rotation, as long as the object stays the same (configuration-wise universal).

#### *2.3.4 Constitutive Relationship*

The constitutive relationship or fluid-structure constitution interchangeably refers to the linking of fluid and structure by Koopmanism's inherent statistical correlations. The eigenfrequency of a Koopman tuple is like its DNA. When it matches that of another on the same, invariant Koopman subspace, the association is deterministic and spectrally warranted.

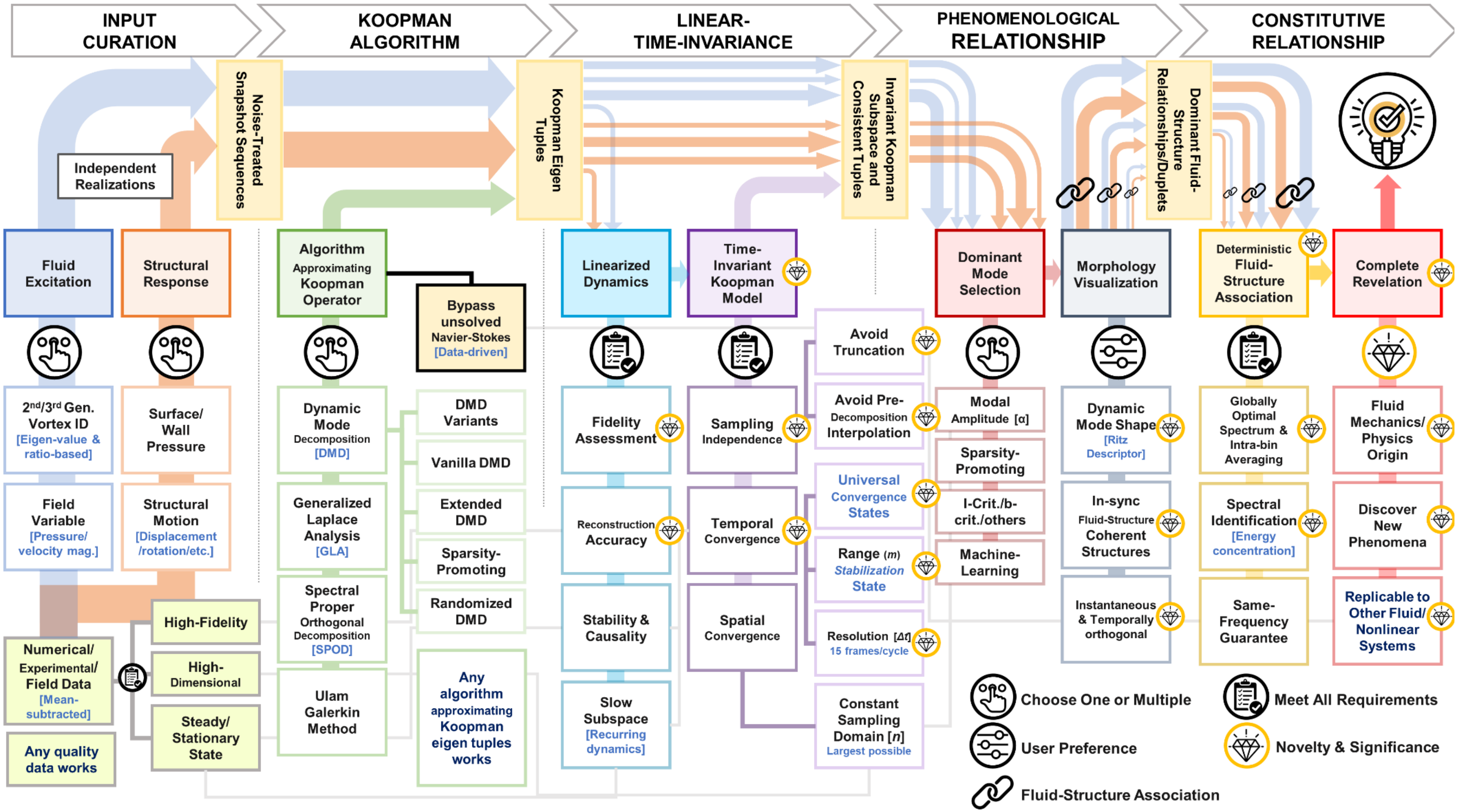
INPUT CURATION
KOOPMAN ALGORITHM
LINEAR-TIME-INVARIANCE
PHENOMENOLOGICAL RELATIONSHIP
CONSTITUTIVE RELATIONSHIP
Noise-Treated Snapshot Sequences
Koopman Eigen Tuples
Invariant Koopman Subspace and Consistent Tuples
Dominant Fluid-Structure Relationships/Duplets
Independent Realizations
Fluid Excitation
Structural Response
Algorithm Approximating Koopman Operator
Bypass unsolved Navier-Stokes [Data-driven]
Linearized Dynamics
Time-Invariant Koopman Model
Dominant Mode Selection
Morphology Visualization
Deterministic Fluid-Structure Association
Complete Revelation
2nd/3rd Gen. Vortex ID [Eigen-value & ratio-based]
Field Variable [Pressure/ velocity mag.]
Surface/ Wall Pressure
Structural Motion [Displacement /rotation/etc.]
Numerical/ Experimental/ Field Data [Mean-subtracted]
Any quality data works
High-Fidelity
High-Dimensional
Steady/ Stationary State
Dynamic Mode Decomposition [DMD]
Generalized Laplace Analysis [GLA]
Spectral Proper Orthogonal Decomposition [SPOD]
Ulam Galerkin Method
DMD Variants
Vanilla DMD
Extended DMD
Sparsity-Promoting
Randomized DMD
Any algorithm approximating Koopman eigen tuples works
Fidelity Assessment
Reconstruction Accuracy
Stability & Causality
Slow Subspace [Recurring dynamics]
Sampling Independence
Temporal Convergence
Spatial Convergence
Avoid Truncation
Avoid Pre-Decomposition Interpolation
Universal Convergence States
Range (m) Stabilization State
Resolution [Δt] 15 frames/cycle
Constant Sampling Domain [n] Largest possible
Modal Amplitude [α]
Sparsity-Promoting
I-Crit./b-crit./others
Machine-Learning
Dynamic Mode Shape [Ritz Descriptor]
In-sync Fluid-Structure Coherent Structures
Instantaneous & Temporally orthogonal
Globally Optimal Spectrum & Intra-bin Averaging
Spectral Identification [Energy concentration]
Same-Frequency Guarantee
Fluid Mechanics/ Physics Origin
Discover New Phenomena
Replicable to Other Fluid/ Nonlinear Systems
Choose One or Multiple
Meet All Requirements
User Preference
Novelty & Significance
Fluid-Structure Association

Figure 1: A schematic illustration of the Koopman Linear-Time-Invariance (Koopman-LTI) Architecture. It consists of the *Input Curation*, *Koopman Algorithm*, *Linearly-Time-Invariance*, *Constitutive Relationship*, and *Phenomenological Relationship* modules. Each module contains several submodules outlining the requirements or options. The Koopman-LTI is data-driven and modular, theoretically accommodating all types of input data and solution algorithms that approximate the Koopman eigen tuples. One may follow this perception and analytical procedure to establish fluid-structure relationships. The in-synch, consistent Koopman modes also bear meaningful implications, enhancing understanding of FSI mechanisms.

As we will shortly demonstrate, LTI models of different fluid and structure measurables have spectral distributions that conform to some astonishing agreements. The *a posteriori* consensus warrants fluid-structure constitution, and transform the complex stochastic process (i.e., inhomogeneous anisotropic turbulence) in the Euclidean space into a simple system in the Fourier space, facilitating some incisive characterizations.

#### *2.3.5 Phenomenological Relationship*

The term phenomenology is used here quite literally as the study of the phenomenon. This model visualizes the coherent structures of dominant fluid-structure duplets, analyzes their in-synch behaviors, and underpins their FSI mechanisms. On this note, our discussions are limited to the phenomenological aspect of fluid mechanics. Nonetheless, as Roshko (1993) once pointed out, "the problem of bluff body flow (our test subject) remain almost entirely in the empirical, descriptive realm of knowledge." Therefore, mode shapes---allegedly describing only relative behaviors----still bear profound insights. We will demonstrate the physical interpretations in Part 2.

# 3. Practical Rendering of the Koopman-LTI (Modules 1-3)

Following the architectural introduction, the upcoming sections present an illustrative, pedagogical demonstration of the Koopman-LTI via a canonical fluid-structure system.

## *3.1 Module 1: Input Curation*

#### *3.1.1 Test Subject*

This work employed the most fundamental yet sufficiently challenging case to assess Koopman-LTI's capacity and encourage intellectual resonance with the broadest audience--the subcritical free-shear prism wake (see figure 2). This paradigmatic configuration is geometrically simplistic, phenomenologically complex (Bai & Alam, 2018; Z. Chen et al., 2020), and we have fair knowledge about its phenomenology, so the fluid mechanics does not overshadow the demonstrative purpose (Bai & Alam, 2018; Z. Chen et al., 2021; Lander et al., 2016; Paidoussis et al., 2010; Rastan et al., 2021). The prism wake is also a popular test subject for fluid principles, for example, behaviors of inhomogeneous anisotropic turbulence (Lander et al., 2016, 2018) and the Kolmogorov hypotheses (Portela et al., 2017).

Furthermore, we selected the subcritical regime with inhomogeneous anisotropic turbulence because success with this realistic stochastic system will tell volumes about Koopman-LTI's generality. The Reynolds Number is $Re=U_{\infty}D/\upsilon=2.2\times10^4$, where $U_{\infty}$ is the free-stream speed, $D$ is the prism side length, and $\upsilon$ is the kinematic viscosity, characterizing a vast neighborhood of phenomenological similitude during the shear layer turbulence transition II (Bai & Alam, 2018). The free-shear family (*i.e.*, jet, mixing layers, wakes, *etc*.) also shares many common features, broadening the applicability.

The infinite spanwise length prevents undesired complications due to the end effects (Z. Chen et al., 2018; White, 2006), and an infinite stiffness reduces the complexity from a bi-directional feedback loop to a mono-directional case---a simplification ubiquitously adopted for external flows in large-scale or civil applications (Z. Chen et al., 2020, 2021; Rodi, 1997; Tse et al., 2020; Xinyue Zhang et al., 2020; Xuelin Zhang et al., 2022).

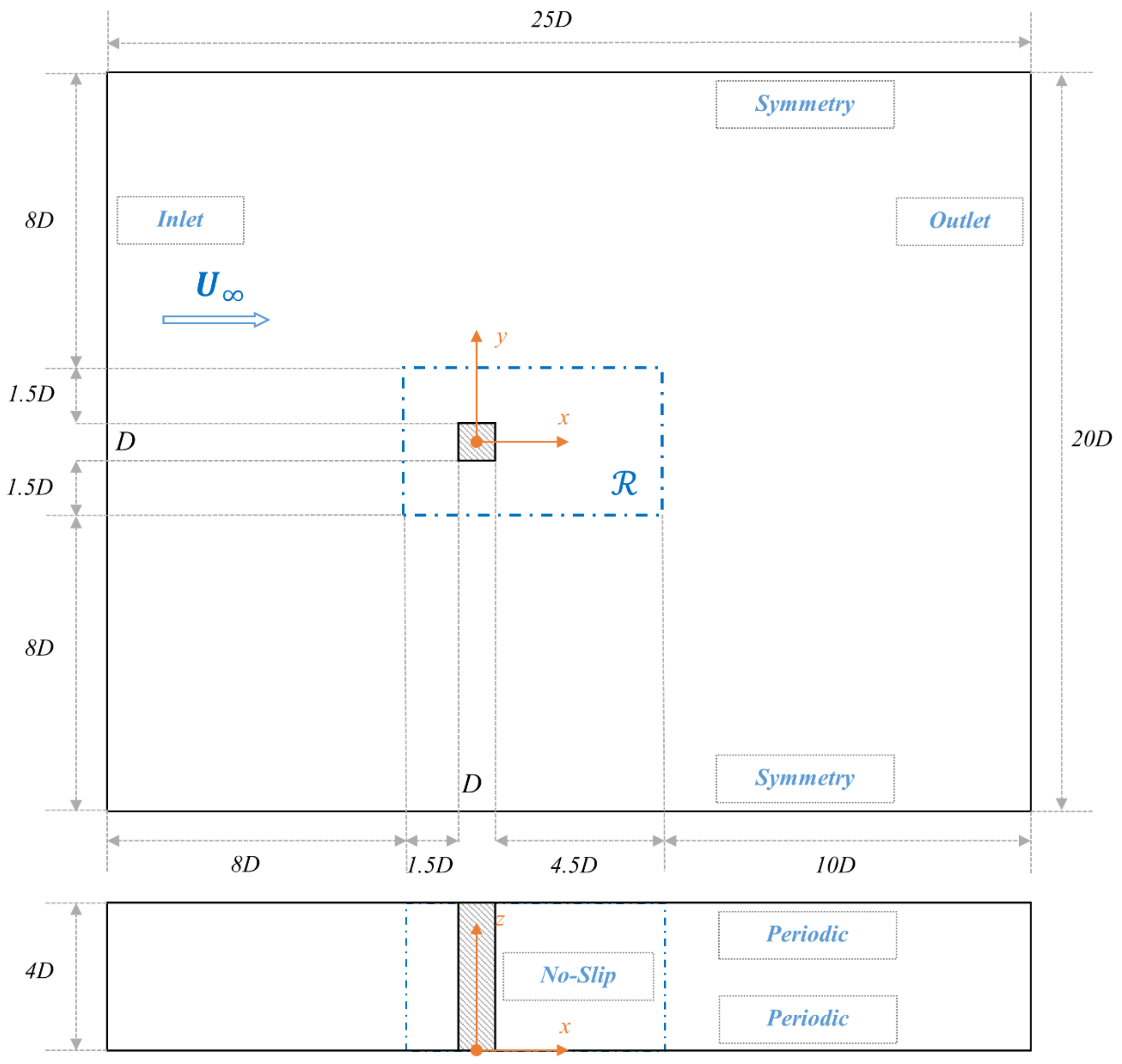

Figure 2: Schematic illustration of the computational domain of the numerical simulation.

### *3.1.2 Numerical Scheme*

High dimensionality is preferred for a better approximation of the infinite-dimensional Koopman operator, while noise is unfavorable, hence the selection of numerical data. The flow is simulated by the Large-Eddy Simulation with Near-Wall Resolution (LES-NWR), as defined by Pope (2000). The simulation adopts the DNS domain from Portela et al. (2017), except using

*4D* for the spanwise length instead of *πD* for computational ease in the Euclidean space (see figure 2).

We employed a finite-volume, segregated, pressure-based solution algorithm for this low-Mach-number incompressible flow. With second-order schemes for spatial and temporal discretizations and a stringent convergence criterion of $O^{-6}$, numerical dissipation and dispersion are minimized. Moreover, the simulation evolved by a non-dimensional time interval $t^*$,

$$t^* = \frac{t\,U_\infty}{D} = 1.61 \times 10^{-3}, \tag{3}$$

where *t* is the physical time step. The Courant-Friedrichs-Lewy criterion CFL<1 is always satisfied, eliminating the time marching issues when solving partial differential equations.

To avoid repetition, we direct the readers to our previous work (C. Y. Li et al., 2022a) for comprehensive simulation details, grid assessment, and case validation. In short, the numerical accuracy is comparable to several DNS renderings (Cao et al., 2020; Portela et al., 2017; Trias et al., 2015).

### 3.1.3 Inventory Measurables

In total, 18 field and wall measurables have been sampled as independent realizations (see table 1). Readers may find the relevant definitions in Li et al. (2022b). The subsequent text refers to the upstream (AB), top (BC), downstream (CD), and bottom (DA) walls according to the orientation in figure 3. After Liu (2019), this work also refers to the vorticity-based vortex identification criterion, namely $|\omega|$, as the first-generation vortex field, the eigenvalue-based criteria, namely $q$ and $\lambda_2$, as the second-generation, and the ratio-based criteria, namely Ω and $\tilde{\Omega}_R$, as the third-generation.

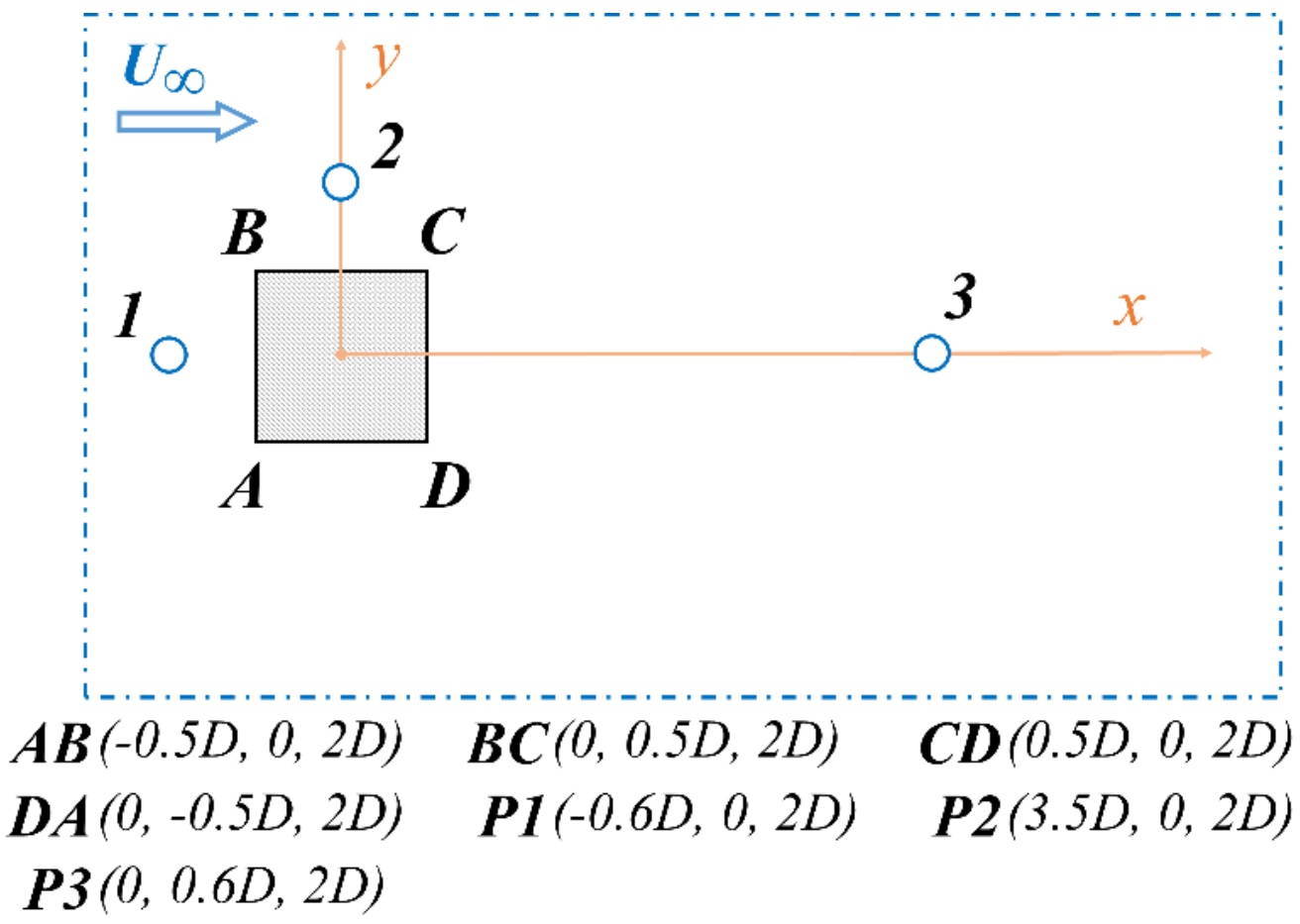

Figure 3: Orientation and location of the prism walls.

| *Wall Pressure* | *Flow Field* | *Turbulence Field* | *Vortex Field* |
|---|---|---|---|
| BC<br>[Top wall] | $P$<br>[Total pressure] | $\langle u'v' \rangle$<br>[Reynolds stress from $u'$ and $v'$] | $\lvert\omega\rvert$<br>[Vorticity magnitude]<br>Helmholtz (1858) |
| DA<br>[Bottom wall] | $u$<br>[x-velocity] | $\langle u'w' \rangle$<br>[Reynolds stress from $u'$ and $w'$] | $q$<br>[$q$-criterion]<br>Hunt et al. (1988) |
| AB<br>[Upstream wall] | $v$<br>[y-velocity] | $\langle v'w' \rangle$<br>[Reynolds stress from $v'$ and $w'$] | $\lambda_2$<br>[$\lambda_2$-criterion]<br>Jeong & Hussain (1995) |
| CD<br>[Downstream wall] | $w$<br>[z-velocity] | $\langle k \rangle$<br>[Turbulence kinetic energy] | $\Omega$<br>[Ω-criterion]<br>Liu et al. (2016) |
| | $\lvert U\rvert$<br>[Velocity magnitude] | | $\tilde{\Omega}_R$<br>[Ω-Liutex criterion]<br>Liu et al. (2018) |

Table 1: Summary of the inventory consisting of 18 measurables.

### *3.2 Module 2: Koopman Algorithm*

This work selected the most ubiquitous Dynamic Mode Decomposition (DMD) among all the aforenoted algorithms to approximate the Koopman eigen tuples. The DMD has many variants to different advantages: the Arnoldi-based formulation, also known as the Koopman Mode Decomposition (Rowley et al., 2009), the companion-matrix formulation (Schmid, 2010), the similarity-matrix formulation (Tu et al., 2014), the sparsity-promoting variant (Jovanović et al., 2014), the extended DMD (Williams et al., 2015), the randomized DMD (Erichson & Donovan, 2016), the recursive DMD (Noack et al., 2016), and so on. We chose the rudimentary similarity-matrix formulation for tractability and robustness with high-dimensional data (Kutz et al., 2016).

Since our focus is not on the algorithm, we layout only the most necessary information to preserve concision. Readers may refer to Tu et al. (2014) or Li et al. (2022b) for the complete formulation.

The DMD deploys a data-driven, finite-dimensional mapping matrix $\boldsymbol{A}$ to approximate the Koopman operator $U$. $\boldsymbol{A}$ connects two time-shifted snapshot sequences, $\boldsymbol{X}_1$ and $\boldsymbol{X}_2$, of a particular measurable

$$\boldsymbol{X}_2 = \boldsymbol{A}\boldsymbol{X}_1. \tag{4}$$

$\boldsymbol{A}$ is an unknown matrix that mimics the map $\boldsymbol{f}$ and Koopman operator $U$. Intuitively, the accuracy of $\boldsymbol{A}$ increases with the dimensionality, so too is the computational expense.

$\boldsymbol{X}_1$ and $\boldsymbol{X}_2$, that are separated by a uniform time step $\Delta t$,

$$\boldsymbol{X}_1 = \{\boldsymbol{x}_1, \boldsymbol{x}_2, \boldsymbol{x}_3, \ldots, \boldsymbol{x}_{m-1}\}, \tag{5}$$

$$\boldsymbol{X_2}=\{\boldsymbol{x_2},\boldsymbol{x_3},\boldsymbol{x_4},\ldots,\boldsymbol{x_m}\}, \tag{6}$$

where $\boldsymbol{x_i} \in \mathbb{C}^n$ are individual data snapshots in the vector form. $n$ denotes the spatial dimension of the data sequence, which is capped by the maximum spatial resolution of a testing apparatus or numerical grid. $m$ denotes the temporal dimension of the data sequence, which is capped by the sample size. Readers are reminded of a tacit assumption of the DMD: $m<<n$.

By a Singular-Value-Decomposition (SVD)-based procedure, one arrives at a similar-matrix $\tilde{\boldsymbol{A}}$ that replaces $\boldsymbol{A}$ in equation (4). What can be done to the $\tilde{\boldsymbol{A}}$ matrix is limited only by the boundaries of linear algebra. This work adopts the default procedure, which is the eigendecomposition.

An eigendecomposition by the Ritz method yields

$$\tilde{\boldsymbol{A}}\boldsymbol{W}=\boldsymbol{W}\boldsymbol{\Lambda}, \tag{7}$$

where $\boldsymbol{W}$ contains the eigenvectors (Ritz vectors) $\boldsymbol{w_j}$, and $\boldsymbol{\Lambda}$ contains the corresponding discrete-time eigenvalues (Ritz values) $\lambda_j$.

The eigen tuples yield the *exact* DMD modes (Tu et al., 2014) as

$$\boldsymbol{\Phi}=\boldsymbol{X_2}\boldsymbol{V}\boldsymbol{\Sigma}^{-1}\boldsymbol{W}, \tag{8}$$

where $\boldsymbol{\Phi}$ contains the Koopman/DMD mode $\boldsymbol{\phi_j}$. $\boldsymbol{\Sigma}$ *and* $\boldsymbol{V}$ are outcomes of the SVD and contains the singular values $\sigma_j$ and temporally orthogonal modes $\boldsymbol{v_j}$, respectively.

Every mode $\boldsymbol{\phi_j}$ also corresponds to a physical frequency $\omega_j$ in continuous time

$$\omega_j=\Im\{log(\lambda_j)\}/\Delta t, \tag{9}$$

and a growth/decay rate $g_j$

$$g_j = \Re\left\{log(\lambda_j)\right\}/\Delta t. \tag{10}$$

This procedure for computing the exact DMD modes is an algorithm for finding the Koopman eigen tuples (Tu et al., 2014), and any incongruence between the modes is attributed to the artifact of the approximated Koopman operator (Rowley et al., 2009). Theoretically, an infinite-dimensional Koopman mode is perennially oscillatory with a zero growth/decay rate.

## *3.3 Module 3: Linear-Time-Invariance*

Module 3 begins the analytical end by ensuring that the algorithm outcomes are accurate, stable, and sampling-independent, capturing all the long-term, recurring dynamics of the input system.

### *3.3.1 Linearity*

Nonlinearity causes possible inequalities between a phenomenon's forward and reverse paths. The directional bias makes fluid-structure constitutions extremely difficult, if not impossible. Module 2 effectively linearizes the nonlinearities and decomposes an input measurable into a linear superposition

$$\boldsymbol{x}_{Koopman,i} = \sum_{j=1}^{r} \boldsymbol{\phi}_j \, exp(\omega_j t_i)\alpha_j. \tag{11}$$

where $\boldsymbol{x}_{Koopman,i}$ are the Koopman eigen tuples of different spatiotemporal weights, which sum into the Koopman reconstruction of the input data at instant *i*. *r* denotes the truncation order of $\tilde{\boldsymbol{A}}$ and $\alpha_j$ denotes the coefficient of weight, or the modal amplitude.

Instead of the conventional static DMD mode shape, we define an evolutionary, dynamic mode shape

$$\boldsymbol{M}_j = \boldsymbol{\phi}_j \, exp(\omega_j \boldsymbol{t}) \alpha_j, \tag{12}$$

in which

$$\boldsymbol{t} = \{\Delta t,\ 2\Delta t, 3\Delta t, \ldots, (m-1)\Delta t\}. \tag{13}$$

$\boldsymbol{M}_j$ contains a fragment of the total information that has been harmonically averaged over a prescribed frequency span (Rowley et al., 2009). They together formulate the globally optimal linearization. It is important to note that $\boldsymbol{M}_j$ is temporally orthogonal.

### 3.3.2 Time-Invariance

Time-invariance is the other critical aspect of the Koopman-LTI. Since the spatial dimension $n$ is controlled by the number of measurement nodes and is typically fixed in practice, sampling independence is essentially a matter of temporal convergence. Our previous work found LTI for the DMD, identified four universal convergence states for pragmatic convergence, and established a best practice for engineering applications (C. Y. Li et al., 2022a). Accordingly, one may fulfill time-invariance by meeting the following conditions.

#### *Mean-Subtraction*

The first step is to supply mean-subtracted input. Although the original Koopman/DMD analysis does not impose the requirement, it is recommended because the mathematical origin of the Koopman analysis traces back to the Discrete Fourier Transform (DFT) or the Z-transform. Chen *et al.* (2012) have formally demonstrated that the Koopman/DMD modes are equivalent to DFT modes for zero-mean data. Rowley et al. (2009) also pointed out that any incongruence between the Koopman/DMD and DFT modes is attributed to the artifact of approximating the Koopman operator. For these reasons, Towne et al. (2018) recommended mean-subtraction in practice to minimize the incongruence.

To understand the requirement, readers are reminded of the rudimentary Z-transform,

$$\mathcal{Z}\{x[k]\} = \sum_{k=0}^{\infty} x[k]z^{-k}, \tag{14}$$

where $k \in \mathbb{Z}^{+}$ yields a unilateral Z-transform and $z=be^{j\theta}$. One may think of the Koopman linearization as a dynamical sweep by sinusoids and exponentials. We found that the mean-field mainly contains non-oscillatory dynamics (C. Y. Li et al., 2022b). Therefore, the fundamental justification for mean-subtraction is to prevent reluctant descriptions of non-oscillatory dynamics by oscillatory descriptors. To this end, all 18 data sequences have been mean-subtracted.

### *Statistical Stationarity*

The second step is to supply steady or statistical stationary data. Although the Koopman analysis does not limit its scope to stationary flows, but as one can imagine, the sinusoidal and exponential sweep is more suited for recurring, oscillatory dynamics. Accordingly, steady or stationary data substantially elevates a Koopman linearization's stability.

Stationarity is also critical to underpinning predominant dynamical contributors. As commented by Williams et al. (2015), on a particular subspace of the Koopman operator, one may capture the long-term dynamics of an input observable in a span of eigenfunctions associated with eigenvalues near the unit circle in discrete time. The span is known as the *slow* subspace. The opposite, the *fast* subspace, captures the transient dynamics that quickly emerge and dissipate. In the case of a fluid system, the substance of the slow subspace translates to steady or stationary phenomena, while the subspace *per se* facilitates a low-dimensional approximation of the Koopman operator, or, equivalently, the Navier-Stokes equations. As such, steady dynamics is also a major interest of engineering applications. Accordingly, the 18 measurables have been sampled in the statistically stationary state.

*Temporal Convergence*

The third step is temporal convergence, which refers to the condition that the mapping matrix $\boldsymbol{A}$ is independent of changes of $\boldsymbol{X_1}$ and $\boldsymbol{X_2}$. In this specific rendering, it refers to $\tilde{\boldsymbol{A}}$ being independent of $m$ and $\Delta t$, or the sampling range and resolution, respectively.

We define the grand mean $l_2$-norm of reconstruction error to assess the temporal convergence,

$$G_{\|e\|_2}=\frac{1}{m}\sum_{i=1}^{m}\|e\|_{2,ins,i}, \tag{15}$$

where $\|e\|_{2,ins}\in\mathbb{R}^+$

$$\|e\|_{2,i}=\|e\|_{2,ins}=\frac{1}{n}\sum_{k=1}^{n}\left[\left(\frac{x_{Koopman,k,i}-x_{k,i}}{x_{k,i}}\right)^2\right]^{1/2}, \tag{16}$$

is the instantaneous, spatially-averaged, and $l_2$-normalized reconstruction error, and $\|e\|_{2,rms,i}\in\mathbb{R}^+$

$$\|e\|_{2,rms,i}=\left[\frac{1}{n}\sum_{k=1}^{n}\left(\frac{x_{Koopman,k,i}-x_{k,i}}{x_{k,i}}\right)^2\right]^{1/2}, \tag{17}$$

is its root-mean-squared (rms) value.

Our serial parametric study discovered four convergence states (see figure 4) by varying the sampling range $m$ (C. Y. Li et al., 2022a, 2022b). The study considered both prism and cylinder wakes and observed the same behaviors. After analyzing the states' spectral implications, it was concluded that temporal convergence corresponds to the resolution sufficiency of a spectrum's discretization by Koopman modes, which appear as discrete frequency bins. Therefore, convergence states are *universal* to all DMD decompositions.

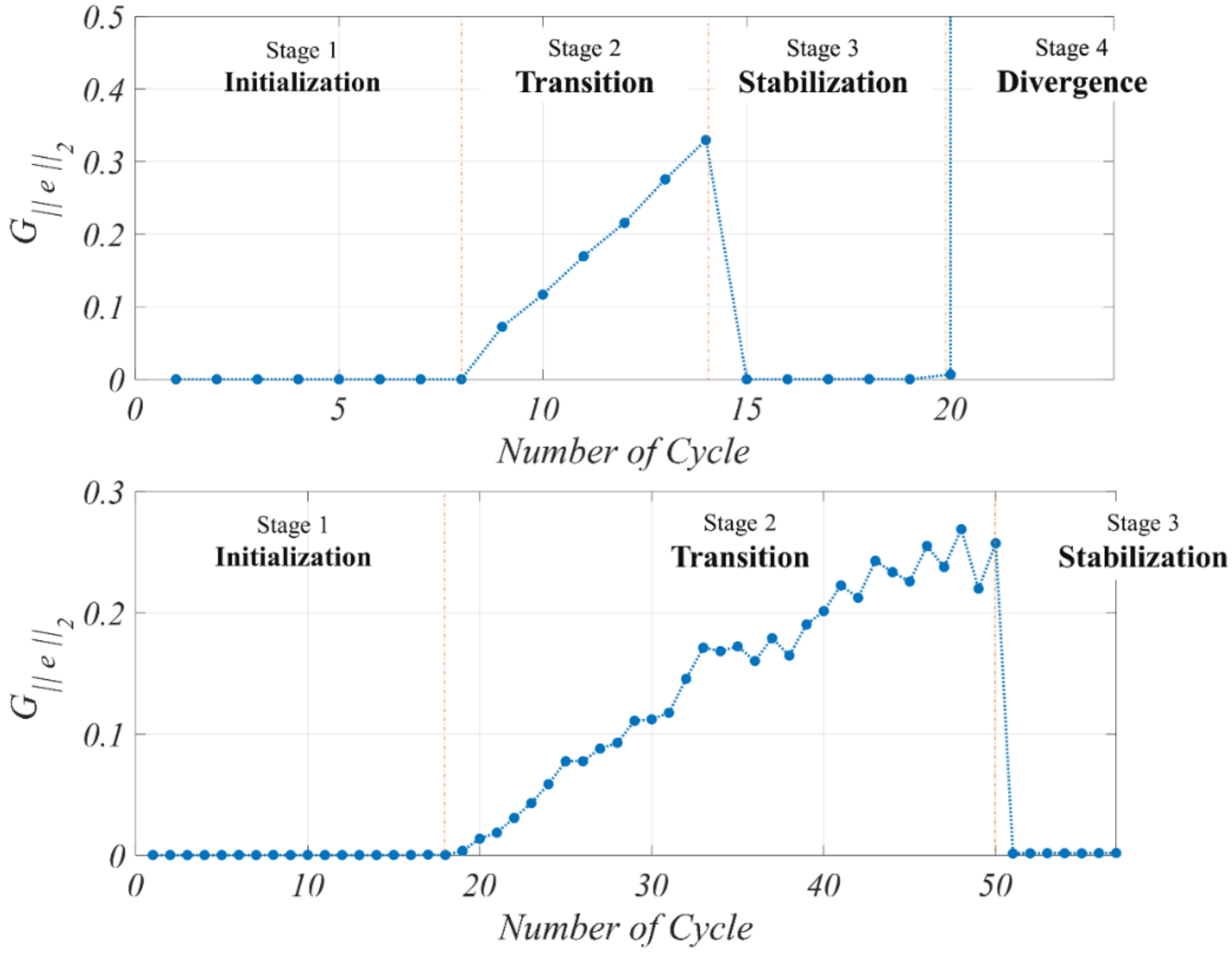

Figure 4: Grand mean error versus sampling resolution with the prism wake (top) and cylinder wake (bottom). Image reproduced with the authorizations of a Creative Commons Attribution 4.0 International License.

Li, C. Y., Chen, Z., Tse, T. K. T., Weerasuriya, A. U., Zhang, X., Fu, Y., & Lin, X., *Nonlinear Dynamics*, 1–25, 10.1007/s11071-021-07167-8, (2022); licensed under a Creative Commons Attribution (CC BY) license.

Li, C. Y., Chen, Z., Tse, T. K. T., Weerasuriya, A. U., Zhang, X., Fu, Y., & Lin, X.,, ArXiv ID: 2110.06577, (2022); licensed under a Creative Commons Attribution (CC BY) license.

To this end, the *Stabilization* state marks sampling independence. In practice, one can plot the grand mean error against the sampling range, and a sudden error drop after an initial accumulation signals the arrival of the *Stabilization* state. Care is needed because the *Initialization* state also appears with trivial errors, but the eigen tuples are highly unstable. In addition to the grand mean error, one shall also examine $\omega_j$ and $g_j$ with respect to $m$. If no apparent changes are observed, the *Stabilization* state is consolidated.

The convergence of sampling resolution $\Delta t$, as well as its relationship with the range $m$, were assessed by another bi-parametric study. The iso-surface of the leading Koopman mode conveys the general message (see figure 5). The convergence conditions of range and resolution are disentangled and can be assessed independently. Specifically, the convergence of resolution indicates a sufficient the upper range for the discrete spectrum and projects only a mode-specific effect. For general practice, one shall resolve the periodicity of the dynamics of interest by no less than 15 frames per cycle.

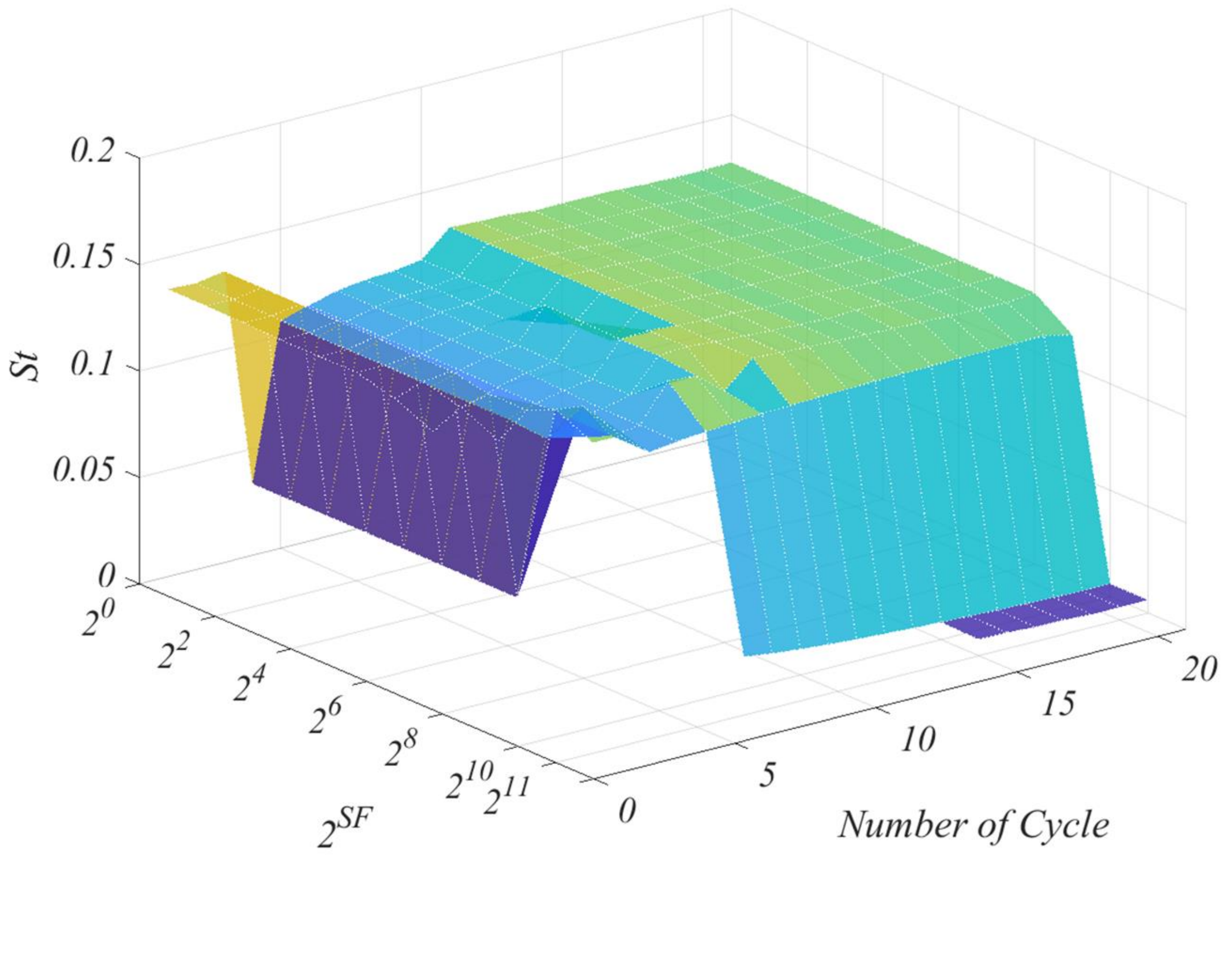

Figure 5: The Strouhal number versus the sampling resolution $2^{SF}$ versus the number of sampled cycles of Mode 1. Image reproduced with the authorizations of a Creative Commons Attribution 4.0 International License.

To avoid repetition, readers may refer to our previous works (C. Y. Li et al., 2022a, 2022b) for more details about temporal convergence. We also disambiguate that the *de jure* time invariance translates to the discrete-continuous transition of the spectral discretization, which is only possible as $m$ approaches infinity and is clearly impractical. Therefore, our discussions here are limited to a pragmatic scope.

### *Truncation and Interpolation*

Users are also advised against truncation because it, even by a single order, notably weakens the stability of the Koopman eigen tuples. the singular-value-based truncation criterion may neglect some low-energy states vital to the overall dynamical content, hampering the modeling fidelity (Noack et al., 2008; Schmid, 2010).

It is also essential to upkeep the $m<n$ condition if one deploys the DMD algorithm because otherwise, the algorithmic outcomes become completely degenerate (C. Y. Li et al., 2022b). High-order pre-decomposition interpolation is recommended to increase $n$ for experimental or field data, which are often inherently low-dimensional in space. However, one shall reckon that interpolation inevitably adds synthetic dynamics into the original system. High-order schemes simply better prevent the entanglement between the artificial and original content.

### 3.3.3 *Accuracy and Stability*

Meeting all the aforenoted requirements, this practical rendering sampled 500 snapshots for each of the 18 measurables in the 3D numerical domain with the inter-snapshot step $\Delta t = 400t^*$. The independent, mean-subtracted realizations were sampled in the stationary state. The range *m* spans 20 oscillations cycles and meets the *Stabilization* state. The resolution $\Delta t$ also resolves the prism wake's predominant shedding cycle at $St=0.127$ by *25* frames per cycle. No truncation and interpolation were performed on the data sequences.

The immediate fruit of the LTI is some astonishing accuracy and stability improvements. Figure 6 presents the instantaneous mean and rms reconstruction error of the 18 LTI systems. The maximum mean and rms errors are $O^{-9}$ and $O^{-6}$, respectively. Excluding the singularities, the maxima further reduce to $O^{-12}$ and $O^{-9}$, which are basically numerical zeros. The virtually non-existent error is extraordinary by any standard: it means the Koopman linearization is infinitely close to exact. Compared to the non-LTI systems generated in our previous efforts (C. Y. Li et al., 2020a, 2020b; Zhou et al., 2021a, 2021b), the improvement is by as much as several orders of magnitude.

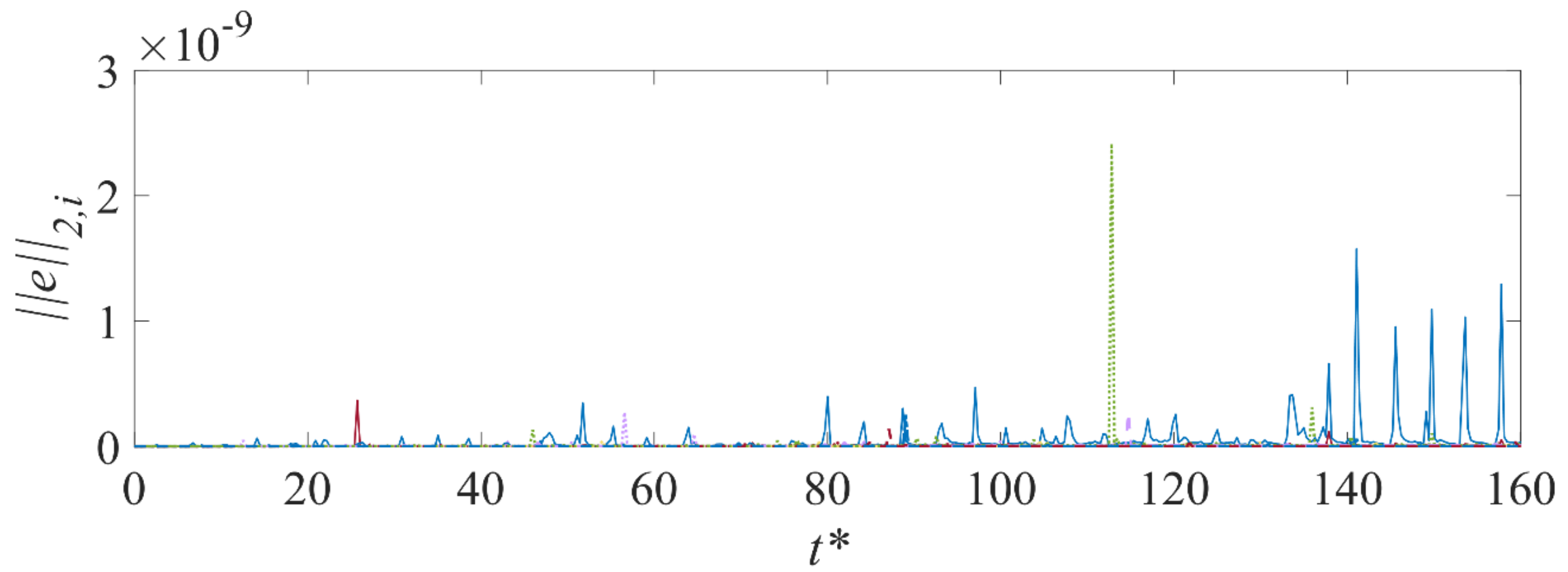

a)

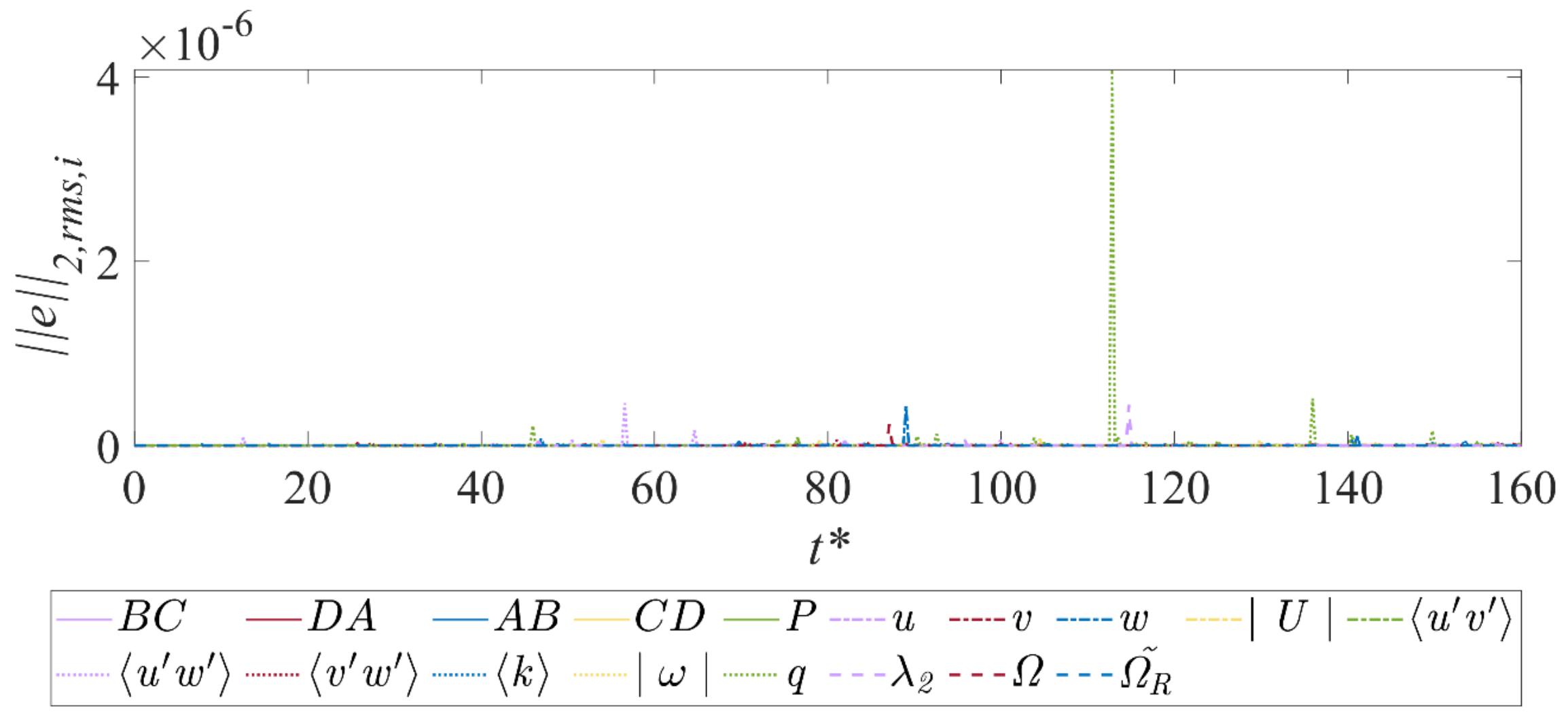

b)

Figure 6: (a) Mean reconstruction error and (b) root-mean-square reconstruction error of the 18 Koopman-LTI systems versus nondimensional time *t**.

On the other hand, stability measures of how well a Koopman system behaves with regularity. It also reflects how well periodic descriptors, the sinusoids and exponentials, depict the input dynamics. The 18 LTI system's Regions of Convergence (ROCs) are presented in figure 7 to assess their stability. For the idealistic, infinite-dimensional Koopman operator *U*, its eigen tuples, or poles, are perfectly oscillatory, perpetually stable, and sit exactly on the $\Re^2+\Im^2=1$ unit circle.

Two key observations are derived from the ROCs. First, all poles lie infinitely close to the unit circle, showing their near-perfect oscillation, hence stellar stability. This attests to the adequacy of the Dynamic Koopman modes to describe input dynamics and, more importantly, the slow subspace. Moreover, the growth/decay rates $g_j$ presented is the numerical index that quantifies stability. A perfect linearization yields $g_j = 0$. The maximum $g_j$ in figure 8 is in the order of $O^{-8}$. Excluding the case of the upstream wall AB, the maximum reduces to $O^{-12}$. Again, the numerical zero testifies for the LTI models' extraordinary stability.

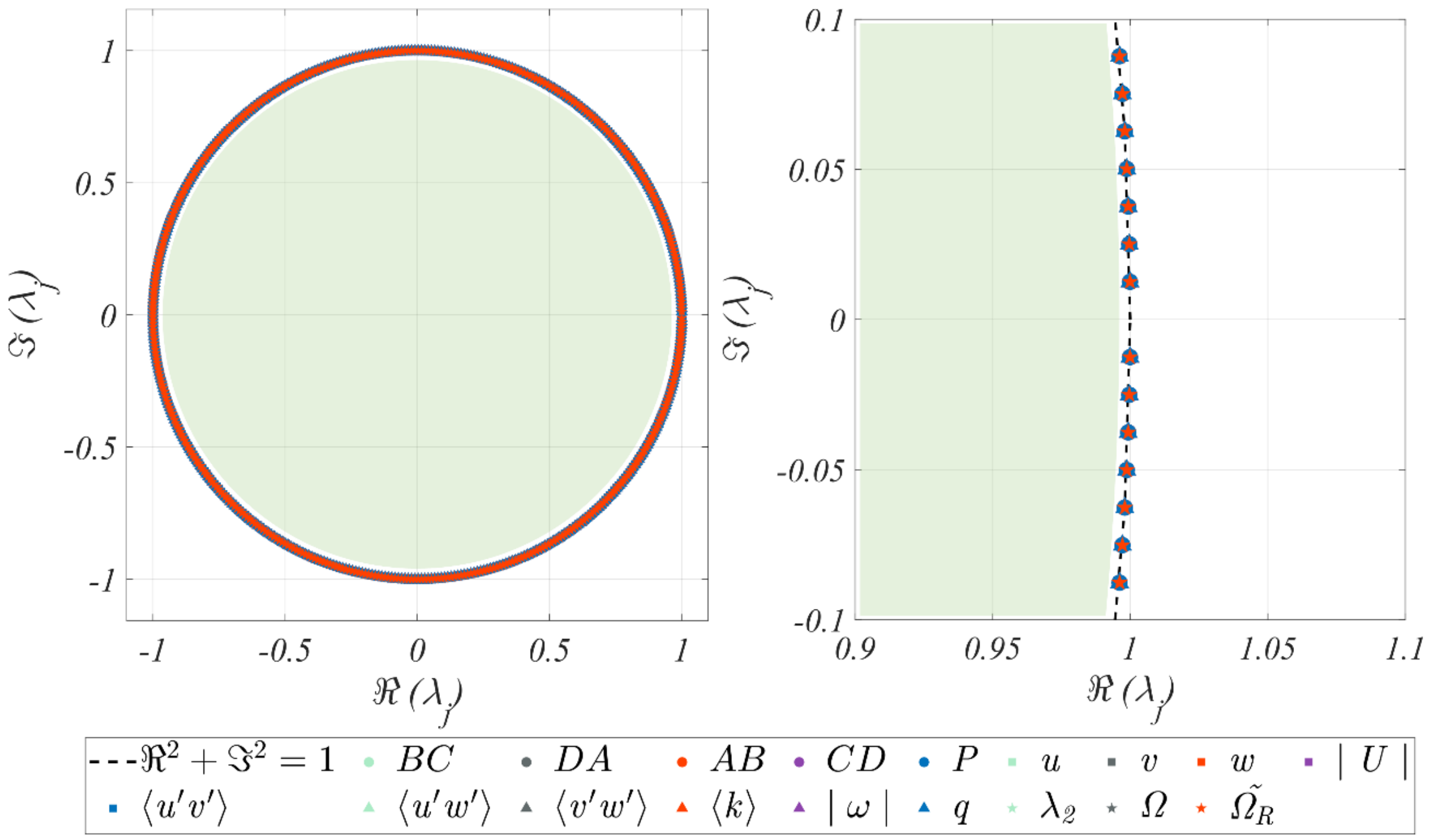

Figure 7: The Region of Convergence of the 18 Koopman-LTI systems (left) and zoomed-in (right) near $\Re(\lambda_j)$=1 and $\Im(\lambda_j)$=0.

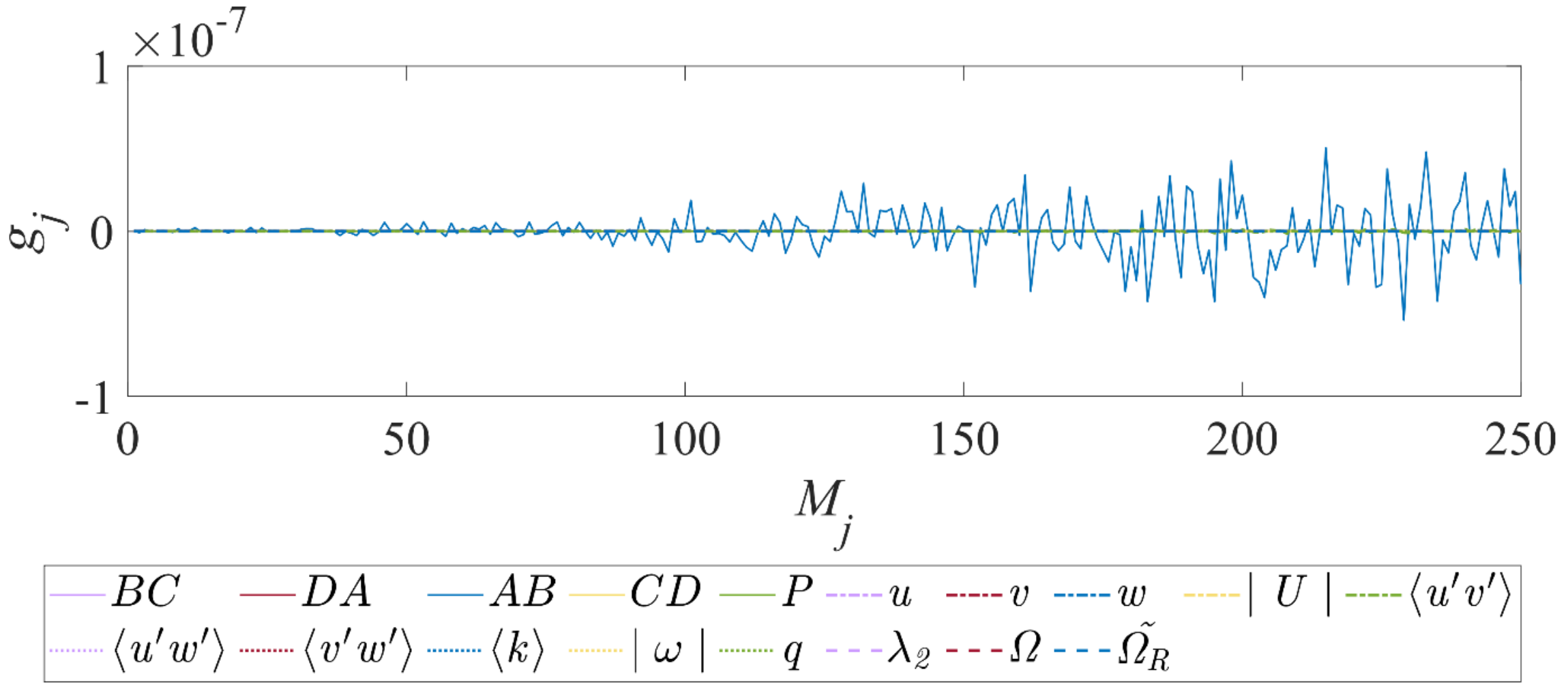

Figure 8: Growth/decay rate of the Koopman-LTI modes of the 18 Koopman-LTI systems versus mode number $\boldsymbol{M}_j$.

The second observation is that the ROCs (in mint green) are characteristic of acausal systems. Acausality guarantees a system's behavior does not depend on past input but only on future ones (Oppenheim et al., 1997). A concomitant interpretation is that the current sample sufficiently captures all the major (or even minor) dynamical contributors, so the entire slow subspace. Again, this is concrete evidence of sampling independence.

At the end of Module 3, it is worth mentioning that Modules 1, 2, and 3 forms an iterative, input-decomposition-assessment feedback loop. The process is analogous to the training process of a neural network, and the grand mean error and growth/decay rates resemble rudimentary residual metrics. This similarity even motivated several Koopman-inspired neural networks (Q. Li et al., 2017; Lusch et al., 2018), which may well serve the role of the Koopman algorithm in Module 2. These extensions are attractive directions for future endeavors. For our particular scope, we reiterate the importance of a sampling-independent model and will shortly demonstrate its vast analytical.

# 4. Constitutive Relationship (Module 4)

Another observation from figure 7 is remarkably inspirational---the Koopman-LTI modes for all 18 systems superpose exactly onto one another. It implies that all systems consist of the same Dynamic Koopman modes $\boldsymbol{M}_j$ or have been discretized into an identical set of eigenfrequencies. This is a pivotal disclosure: the LTI distributes input dynamics across a uniform set of equidistant, discrete frequency bins, over which the temporal content is orthogonal, and the spatial content is bin-wise averaged.

We emphasize the fundamental awareness that no matter how unsolvable or even indiscernible the governing dynamics are, fluid and structure must conform to some overarching laws and physics deeply embedded in the data. Extending this thought, if the input measurables, through which the LTI models are generated, appropriately represent the fluid and the structure, then the extracted eigen tuples quintessentially represent the linearized, orthogonally disentangled principles of FSI. Accordingly, one may exploit the Koopman-LTI to deterministically associate a structural response with its corresponding fluid excitation by frequency-matching, forming the constitutive relationship. Module 4 demonstrates the realization of this perception.

### *4.1 Intra-Group Dynamics*

Although the frequency bins $St_j$ are universal, the coefficients of weight $\alpha_j$ are vastly different, implying disparity in the measurables' spatiotemporal content. Our intention to assess the disparity fostered the definition of the normalized modal amplitude, as

$$-1 \leq \left|\tilde{\alpha}_j\right| \in \mathbb{R} \leq 1. \tag{18}$$

The ranking of $\boldsymbol{M}_j$ in their respective Koopman-LTI systems directly reflects the spatiotemporal disparity, as summarized in table 2. Besides $\alpha_j$, readers are also reminded of several other criteria (Jovanović et al., 2014; Kou & Zhang, 2017; Sayadi et al., 2014), each to their pros and cons, for user selection. Selecting the ten most dominant eigen tuples for each measurable (highlighted in table 2) results in precisely 30 across the entire inventory. The upcoming analysis categorizes the measurables by their origin and assesses their intra-group dynamics. For each measurable, $|\tilde{\alpha}_1| = 1$ for the most dominant tuple.

#### *4.1.1 Prism Walls*

As shown below, figure 8 presents the $|\tilde{\alpha}_j|$ versus $St_j$ spectra and the discrete Koopman modes of the four wall measurables. $|\tilde{\alpha}_j|$ of the top (BC), bottom (DA), and upstream (AB) walls, collectively referred to as the *on-wind walls*, share a remarkable similarity. The most dominant energy concentration, or *peak*, appears at $St=0.1242$. Its broadband content also spreads across several frequency bins in the neighborhood $St=0.1\text{-}0.15$. A secondary narrowband peak resides at *St=0.0497* and has ~*25%*$|\tilde{\alpha}_1|$. Other peaks are also faintly visible, for example, at *St=0.*2422 with ~*10%*$|\tilde{\alpha}_1|$, but are deemed trivial under the overwhelming first two. For clarification, each peak on the spectrum corresponds to a natural flow structure (Hussain, 1986), and $|\tilde{\alpha}_j|$ allegorically represents the energy associated with it. Equivalently speaking, the peak amplitude inversely represents the energy required to excite a natural structure---the greater the $|\tilde{\alpha}_j|$, the more natural the structure.

In contrast to the similitude of the on-wind walls, the downstream wall (CD) exhibits a fundamentally different distribution. Although the most dominant broadband peak still appears at *St=0.1242*, the secondary peak at *St=0.0497* is buried. Instead, several other peaks, at *St=0.0683*, *0.0745*, *0.1739*, *0.1925*, *0.2422*, and *0.3664* with *37.6%*, *42.0%*, *55.6%*, *40.6%*, *32.6%*, and *37.2%*$|\tilde{\alpha}_1|$, respectively, overtake its dominance. An interpretation suggests more complications underlie the downstream wall, making it utterly different from the on-wind walls.

Moreover, the difference between the primary $|\tilde{\alpha}_1|$ and other $|\tilde{\alpha}_j|$ is marked reduced from ~*75%* to *44.5%*, implying the interwoven physics' undermining of the predominant mechanism. The observations allude to the findings of several works on the negative base pressure, which results from the tumultuous vortex activities and entrainment in both the stream- and span-wise directions inside a turbulent wake (Lander et al., 2016; Luo et al., 1994; Unal & Rockwell, 1988). Based on this analysis, we refer to *St=0.1242* as the *primary peak, St=0.0497* as the *secondary peak*, and all others as the *ancillary peaks* in the subsequent text.

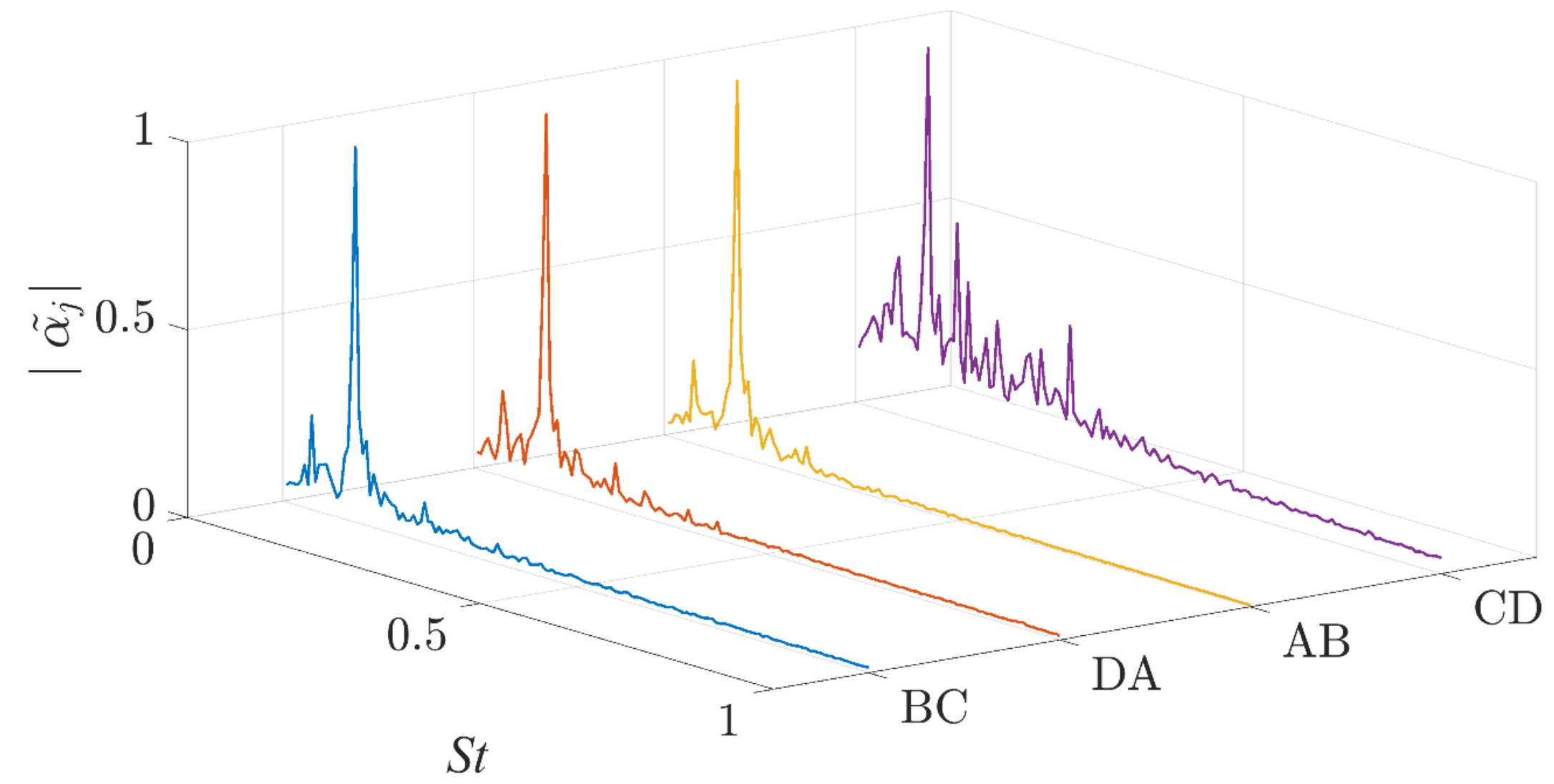

(a)

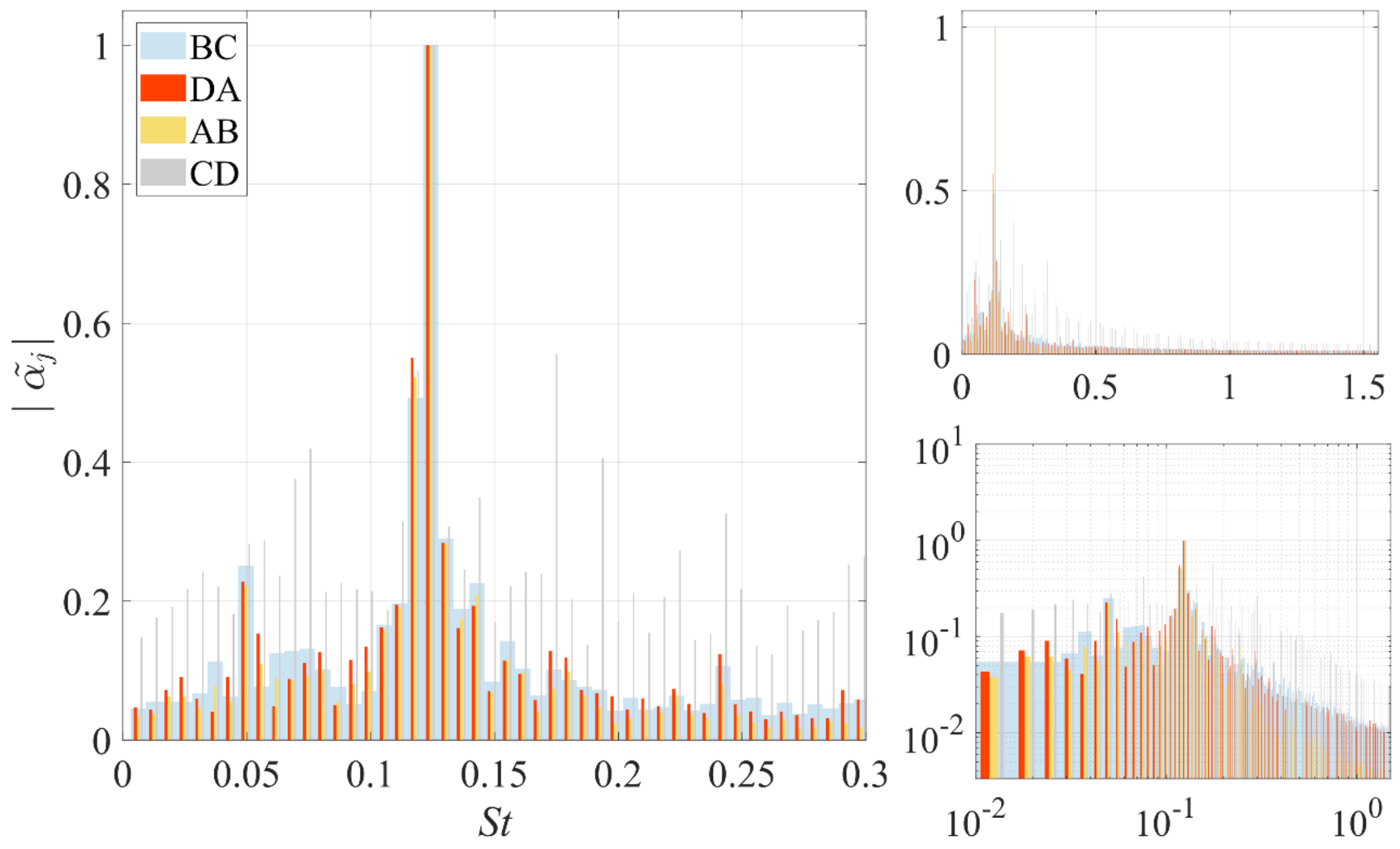

(b)

Figure 9: (a) $|\tilde{\alpha}_j|$ versus *St* of the wall measurables (BC, DA, AB, CD); (b) discrete Koopman modes in linear scale and *St=0-0.3* (left), linear scale and *St=0-1.5* (right top), and log scale and *St=0-1.5* (right bottom).

| ***Mode*** - *Frequency* | **Input Data** | | | | | | | | | | | | | | | | | |
|---|---|---|---|---|---|---|---|---|---|---|---|---|---|---|---|---|---|---|
| | **Primary Measurable** | | | | | | | | | **Turbulence** | | | | **Vortex Identification Criteria** | | | | |
| | Wall *(P)* | | | | Flow Field | | | | | Reynolds Stress | | | TKE | 1st Gen. | 2nd Gen. | | 3rd Gen. | |
| | BC | DA | AB | CD | $P$ | $u$ | $v$ | $w$ | $\lvert U \rvert$ | $\langle u'v' \rangle$ | $\langle u'w' \rangle$ | $\langle v'w' \rangle$ | $\langle k \rangle$ | $\lvert \omega \rvert$ | $q$ | $\lambda_2$ | $\Omega$ | $\tilde{\Omega}_R$ |
| $M_1$ - $St_1$=0.1242 | 1 | 1 | 1 | 1 | 1 | 1 | 1 | 26 | 1 | 11 | 20 | 19 | 1 | 1 | 1 | 1 | 1 | 1 |
| $M_2$ - $St_2$=0.1180 | 2 | 2 | 2 | 3 | 2 | 2 | 2 | 11 | 2 | 14 | 19 | 20 | 3 | 2 | 4 | 4 | 3 | 3 |
| $M_3$ - $St_3$=0.2422 | 14 | 13 | 19 | 9 | 5 | 3 | 5 | 42 | 4 | 19 | 40 | 41 | 10 | 4 | 2 | 2 | 2 | 2 |
| $M_4$ - $St_4$=0.1304 | 3 | 3 | 3 | 11 | 3 | 4 | 3 | 31 | 5 | 23 | 21 | 23 | 5 | 8 | 6 | 7 | 5 | 7 |
| $M_5$ - $St_5$=0.0497 | 4 | 4 | 4 | 14 | 4 | 5 | 8 | 1 | 3 | 5 | 8 | 8 | 2 | 3 | 5 | 5 | 7 | 5 |
| $M_6$ - $St_6$=0.0745 | 10 | 17 | 15 | 4 | 8 | 6 | 13 | 4 | 6 | 13 | 12 | 12 | 19 | 5 | 15 | 16 | 14 | 10 |
| $M_7$- $St_7$=0.0683 | 11 | 21 | 17 | 6 | 10 | 7 | 12 | 14 | 8 | 10 | 11 | 11 | 31 | 7 | 8 | 8 | 11 | 9 |
| $M_8$ - $St_8$=0.1428 | 5 | 6 | 5 | 8 | 6 | 8 | 4 | 6 | 7 | 27 | 23 | 22 | 4 | 9 | 17 | 18 | 8 | 11 |
| $M_9$ - $St_9$=0.1739 | 17 | 11 | 21 | 2 | 9 | 9 | 11 | 24 | 11 | 25 | 28 | 27 | 7 | 6 | 3 | 3 | 10 | 8 |
| $M_{10}$ - $St_{10}$=0.1118 | 6 | 5 | 6 | 10 | 7 | 10 | 6 | 25 | 10 | 21 | 18 | 18 | 6 | 16 | 20 | 22 | 13 | 18 |
| $M_{11}$ - $St_{11}$=0.1366 | 7 | 8 | 7 | 18 | 11 | 11 | 7 | 21 | 9 | 24 | 22 | 21 | 23 | 15 | 25 | 28 | 15 | 19 |
| $M_{12}$ - $St_{12}$=0.1056 | 8 | 7 | 8 | 38 | 12 | 13 | 10 | 22 | 13 | 20 | 17 | 17 | 27 | 17 | 31 | 30 | 19 | 24 |
| $M_{13}$ - $St_{13}$=0.1925 | 23 | 27 | 29 | 5 | 13 | 15 | 9 | 28 | 15 | 22 | 31 | 31 | 11 | 11 | 7 | 6 | 6 | 6 |
| $M_{14}$ - $St_{14}$=0.1553 | 9 | 16 | 9 | 24 | 16 | 16 | 20 | 17 | 21 | 26 | 25 | 26 | 21 | 18 | 10 | 11 | 29 | 25 |
| $M_{15}$ - $St_{15}$=0.0559 | 20 | 9 | 10 | 12 | 15 | 19 | 28 | 23 | 16 | 9 | 9 | 9 | 17 | 10 | 26 | 25 | 25 | 21 |
| $M_{16}$ - $St_{16}$=0.3664 | 37 | 33 | 46 | 7 | 25 | 34 | 15 | 64 | 27 | 45 | 59 | 60 | 16 | 39 | 29 | 26 | 23 | 15 |
| $M_{17}$ - $St_{17}$=0.0994 | 24 | 10 | 13 | 29 | 17 | 14 | 21 | 18 | 17 | 18 | 16 | 16 | 12 | 20 | 37 | 37 | 21 | 20 |
| $M_{18}$ - $St_{18}$=0.0373 | 13 | 44 | 20 | 25 | 24 | 20 | 24 | 2 | 14 | 7 | 6 | 6 | 9 | 19 | 34 | 34 | 26 | 29 |
| $M_{19}$ - $St_{19}$=0.0311 | 25 | 30 | 30 | 19 | 29 | 21 | 32 | 3 | 19 | 6 | 5 | 5 | 14 | 13 | 36 | 36 | 40 | 35 |
| $M_{20}$ - $St_{20}$=0.1677 | 26 | 32 | 32 | 21 | 33 | 25 | 22 | 5 | 32 | 28 | 27 | 28 | 32 | 27 | 60 | 58 | 34 | 28 |
| $M_{21}$ - $St_{21}$=0.0807 | 16 | 12 | 11 | 30 | 14 | 12 | 16 | 7 | 12 | 15 | 13 | 13 | 8 | 12 | 16 | 15 | 16 | 14 |
| $M_{22}$ - $St_{22}$=0.0248 | 35 | 20 | 26 | 27 | 30 | 26 | 42 | 8 | 23 | 4 | 4 | 4 | 22 | 21 | 48 | 48 | 41 | 34 |
| $M_{23}$ - $St_{23}$=0.0124 | 33 | 42 | 34 | 43 | 42 | 32 | 41 | 9 | 31 | 2 | 2 | 2 | 26 | 29 | 58 | 63 | 37 | 36 |
| $M_{24}$ - $St_{24}$=0.0435 | 28 | 19 | 27 | 41 | 26 | 17 | 27 | 10 | 20 | 8 | 7 | 7 | 20 | 14 | 23 | 24 | 18 | 17 |
| $M_{25}$ - $St_{25}$=0.0062 | 45 | 39 | 36 | 53 | 43 | 29 | 30 | 12 | 29 | 1 | 1 | 1 | 28 | 37 | 63 | 64 | 38 | 42 |
| $M_{26}$ - $St_{26}$=0.0186 | 34 | 24 | 25 | 36 | 35 | 31 | 34 | 13 | 28 | 3 | 3 | 3 | 25 | 28 | 49 | 52 | 36 | 39 |
| $M_{27}$ - $St_{27}$=0.0621 | 12 | 38 | 16 | 22 | 19 | 22 | 26 | 20 | 22 | 12 | 10 | 10 | 13 | 22 | 30 | 29 | 31 | 31 |
| $M_{28}$ - $St_{28}$=0.4161 | 50 | 40 | 55 | 45 | 52 | 53 | 56 | 67 | 52 | 62 | 66 | 68 | 53 | 43 | 9 | 9 | 54 | 51 |
| $M_{29}$ - $St_{29}$=0.2236 | 27 | 22 | 23 | 15 | 23 | 30 | 14 | 36 | 33 | 37 | 36 | 37 | 34 | 31 | 12 | 10 | 9 | 12 |
| $M_{30}$ – $St_{30}$=0.2484 | 32 | 35 | 37 | 26 | 22 | 27 | 25 | 40 | 25 | 29 | 39 | 39 | 18 | 32 | 22 | 21 | 4 | 4 |

Table 2: Summary of 30 dominant modes and their respective $|\tilde{\alpha}_j|$ ranking in each Koopman-LTI system (Highlighted: 10 most dominant).

### *4.1.2 Flow field*

We proceed to the field measurables |*U*|, *P*, *u*, *v*, and *w* (see figure 10). Except for *w*, the distributions of |*U*|, *P*, *u*, and *v* exhibit a remarkable resemblance to the on-wind walls, especially the predominance of the primary peak $|\tilde{\alpha}_1|$and the secondary peaks ~*30%*$|\tilde{\alpha}_1|$. The ancillary peaks of considerably higher $|\tilde{\alpha}_j|$ also appear at *St=0.0683*, *0.0745*, *0.1739*, and *0.1925*, increasing from ~*8-13%*$|\tilde{\alpha}_1|$ to ~*15-30%*$|\tilde{\alpha}_1|$ herein. *St=0.2422* even increased from ~*10%* to ~*25%*$|\tilde{\alpha}_1|$. The consistency foreshadows the fluid-structure constitution, which will be demonstrated in section 4.2.

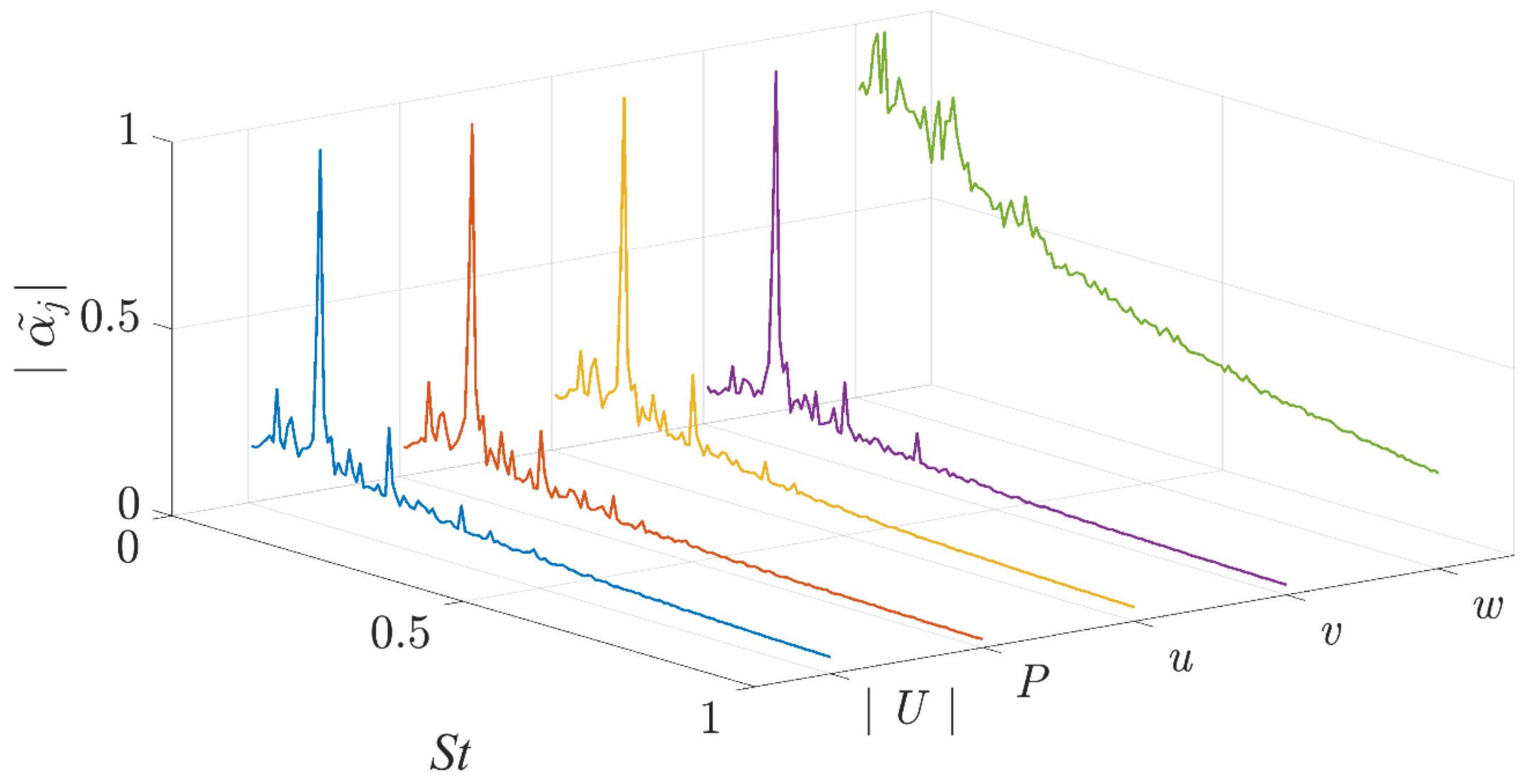

(a)

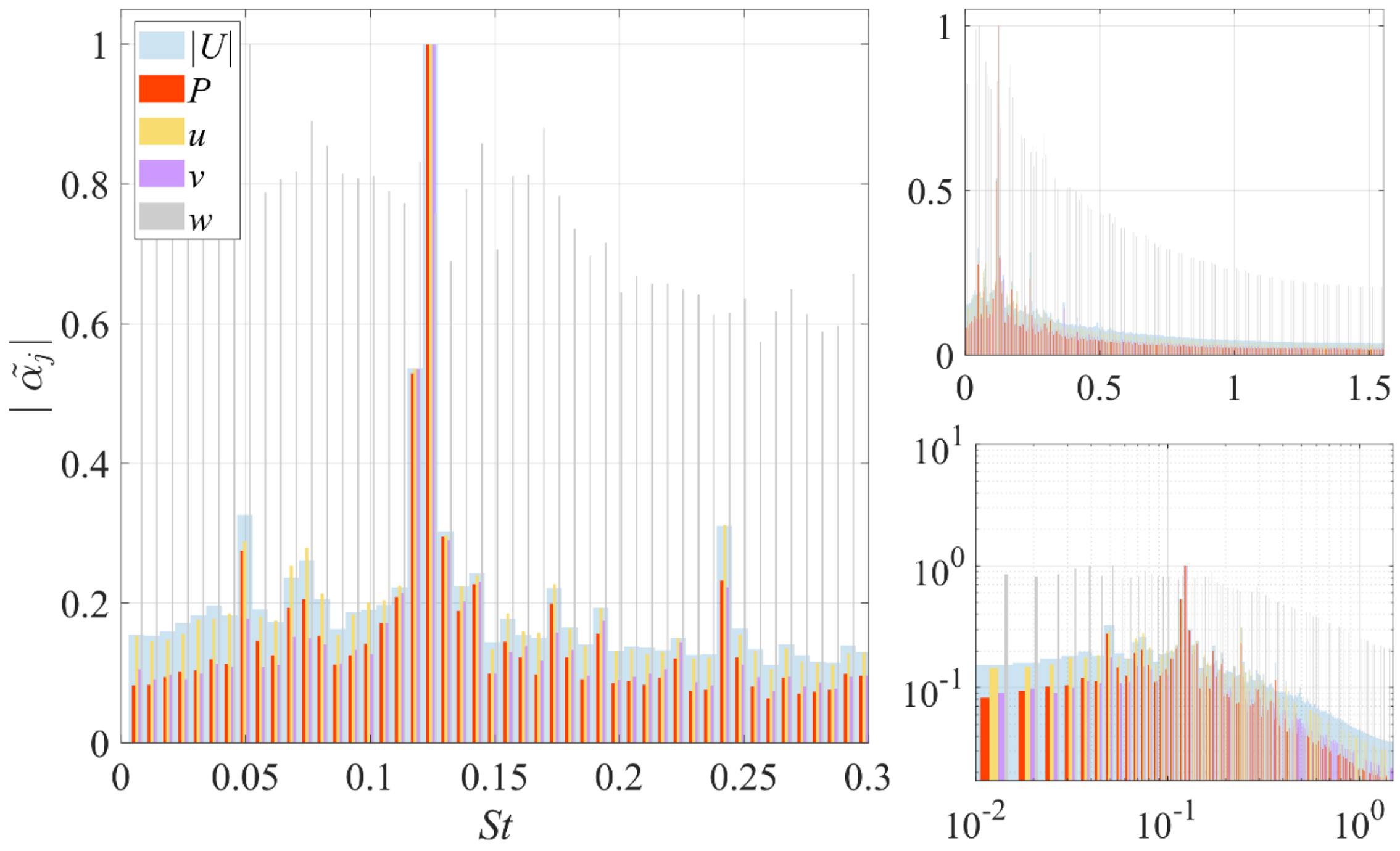

(b)

Figure 10: (a) $|\tilde{\alpha}_j|$ versus *St* of the primary field measurables (|*U*|, *P*, *u*, *v*, *w*); (b) discrete Koopman modes in linear scale and *St=0-0.*3 (left), linear scale and *St=0-1.5* (right top), and log scale and *St=0-1.5* (right bottom).

Continuing with intra-group dynamics, among the three velocity components, the distributions of *u* and *v* are comparable to that of |*U*|, though the resemblance of the former is better than the latter. The observation is attributed to the convection-dominance of free-shear flows (Pope, 2000). Being in alignment with the free-stream, *u* best embodies convection and overwhelms the substance of |*U*|. For this, even with *w* displaying a fundamentally different distribution, |*U*| remains insensitive to *w* and akin to the distribution of *u*.

On a different note, *w* is a peculiarity. Though still visible, the dominance of the primary peak *St=0.1242* is much less prevailing compared to its peers. The *z*-component velocity also weakly reflects the secondary and ancillary peaks. It is to say *w* only marginally contribute to the structure's dominant reactions, or it is dynamically trivial. The observation is lucid on the broadband spectrum (*St=0-1.5)*, particularly in the log scale. Peaks are buried amidst an overarching trend that allegorically appeals to the wavenumber spectrum of the Richardson-Kolmogorov notion (Pope, 2000). To this end, the bleak presence of *w* justifies the selection of several previous efforts, which had studied the three-dimensional prism wake by their planar counterpart (Braza et al., 2006; Ong & Wallace, 1996).

### 4.1.3 Reynolds stresses

The Reynolds stresses display no resemblance to those of the prism walls nor the flow field, but a remarkable consistency between themselves (see figure 11). $|\tilde{\alpha}_j|$ unanimously displays an inverse proportionality with $St_j$ in a largely steady exponential decay. To rationalize the observation, readers are reminded of the implication of the Reynolds stresses---they are the averaged deviatoric components of the stress tensor that accounts for the turbulent fluctuations of fluid momentum. Conceptually, they constitute a measure of an infinitesimal fluid element's distortion, therefore an index quantifying its tendency toward turbulence, or inversely, laminarization.

This knowledge motivates a critical revelation. The spatiotemporal content of the Reynolds stresses metaphorically depicts the eddies. The largest eddies, corresponding to the smallest wavenumbers, extract kinetic energy from the mean-field and channel it into turbulence through production processes. They are, however, the most unstable and vulnerable to distortion. Their highest tendency toward turbulence grants them the greatest spatiotemporal dominance. Although the nonlinear inter-scale transfer can be both forward and inverse, the overall energy balance is negative (Pope, 2000; Portela et al., 2017). So, as the large, more energetic eddies break down into smaller, more abundant ones, $|\tilde{\alpha}_j|$ decays exponentially.

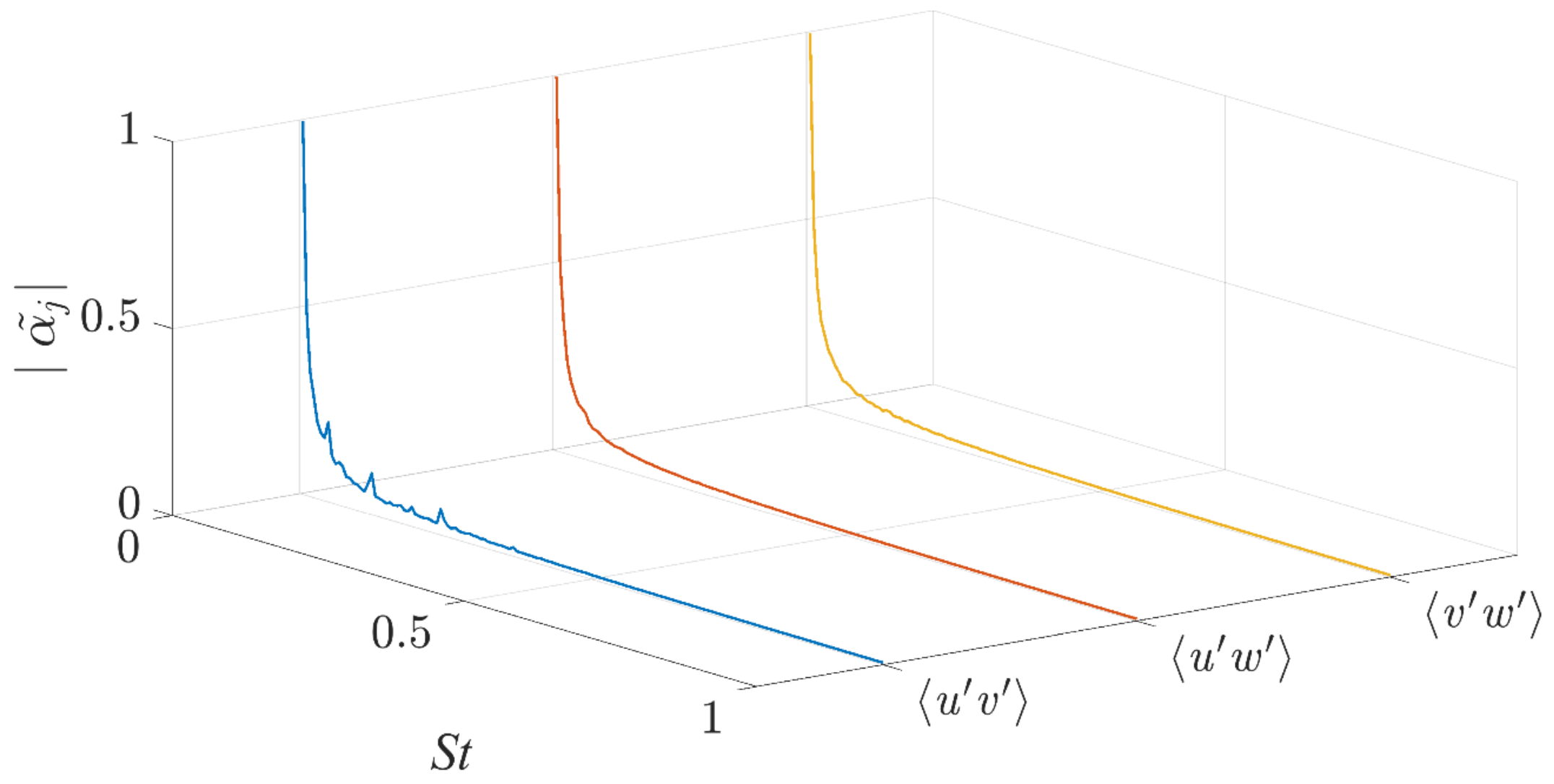

(a)

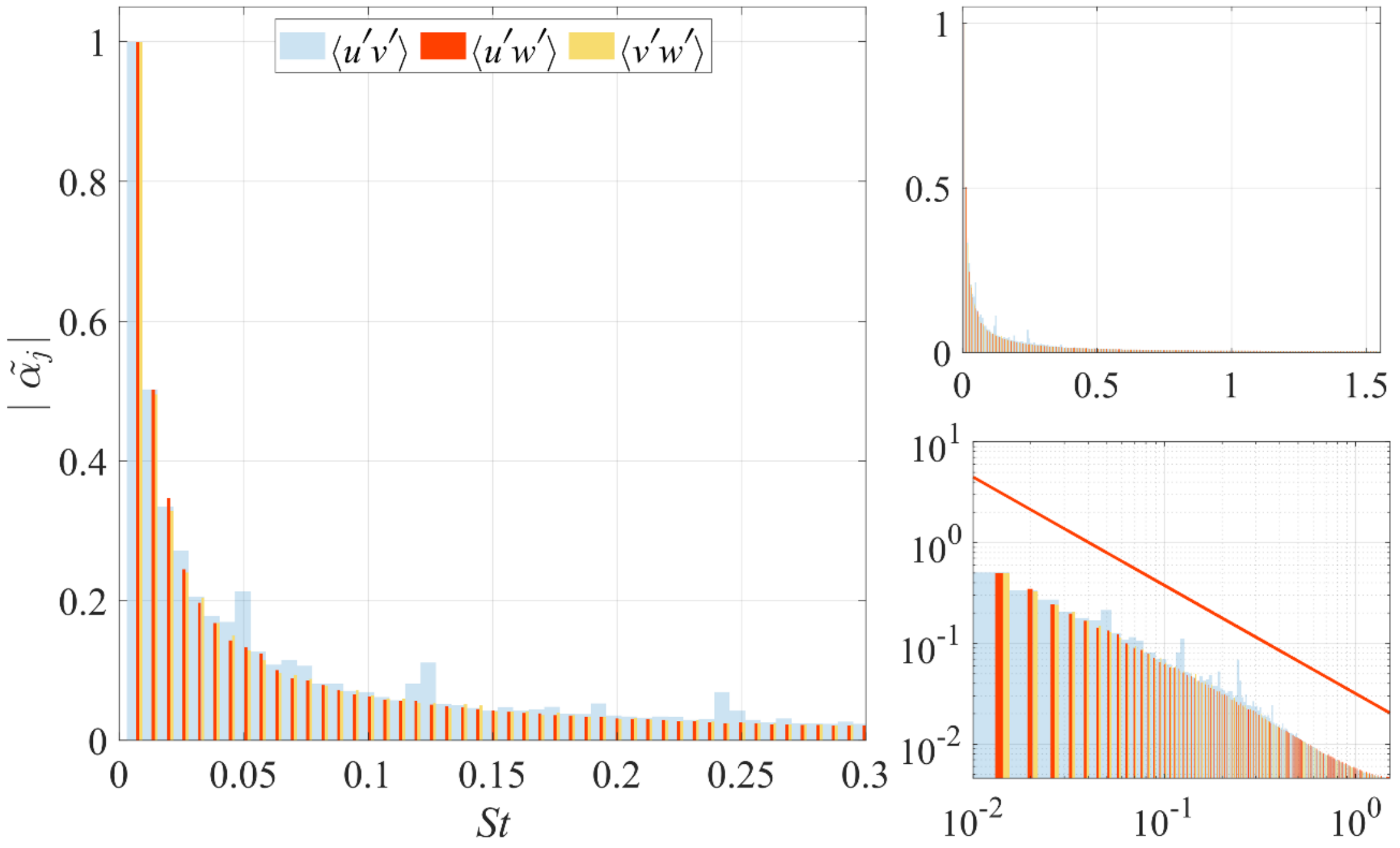

(b)

Figure 11: (a) $|\tilde{\alpha}_j|$ versus *St* of the Reynold stresses (⟨*u'v'*⟩, ⟨*u'w'*⟩,⟨*v'w'*⟩); (b) discrete Koopman modes in linear scale and *St=0-0.3* (left), linear scale and *St=0-1.5* (right top), and log scale and *St=0-1.5* (right bottom), best curve gradient -1.059.

However, at least from a statistically perspective, this depiction is not exactly the cascade nor the inter-scale transfer in the inertial subrange. A best-fit linear regression yields

$$\left|\tilde{\alpha}_{j,\langle u'v'\rangle}\right| = e^{c\,St}, \tag{19}$$

where *c=-1.059, 0.997*, and *1.008* for ⟨*u'v'*⟩, ⟨*u'w'*⟩, and ⟨*v'w'*⟩, respectively. To our best knowledge, *c* does not appeal to any known tenet (e.g., *-5/3*, *4/5*, *4/3*). Further explorations of this observation deserve a dedicated investigative scope. Interestingly, amidst the overwhelming trend, only peaks in ⟨*u'v'*⟩, though far less standout, suggest the presence of the dominant excitations.

#### *4.1.4 Vortex Fields*

The turbulence kinetic energy ⟨*k*⟩ displays more similarity with the vortex fields than the Reynolds stresses (figure 11). ⟨*k*⟩ is the mean kinetic energy per unit mass of eddies derived from the stress tensor's isotropic components (Kundu, 2004; Pope, 2000; White, 2006). So, the empirical observation here is vastly interesting: the dilatory stress components resemble vortex fields more than their deviatoric cousins. Conversely, vortex dynamics are driven by dilation rather than distortion. On this note, the prominent peak at *St= 0.1242* of ⟨*k*⟩ is less acute, and the ancillary ones are *15-30%* more energetic than those of the prism walls and flow field.

The first- and second-generation vortex fields are energetically weaker than ⟨*k*⟩, while the third-generation is more. An emanating remark is that the dynamics of the third-generation criteria are more entwined than those of the others, or its primary peak is less prevailing than its peers. This may have to do with its ratio-based definition (C. Liu, 2019). Another interesting observation is, except for $|\omega|$, all other measurables locate the second peak at *St=0.2422* instead of *St=0.0497*. Inspection reveals this is driven by *u* (figure 9) or the stream-wise convection.

In addition, all vortex fields exhibit mild peaks at *St=0.0683* and *0.0745*, while they are particularly distinctive in ⟨*k*⟩. Finally, despite minor differences, the vortex and flow field measurables generally agree with spectral locations of the energy concentration, namely at *St=0.0497*, *0.0683*, *0.0745*, *0.1242*, *0.1739*, *0.1925*, and *0.2422*.

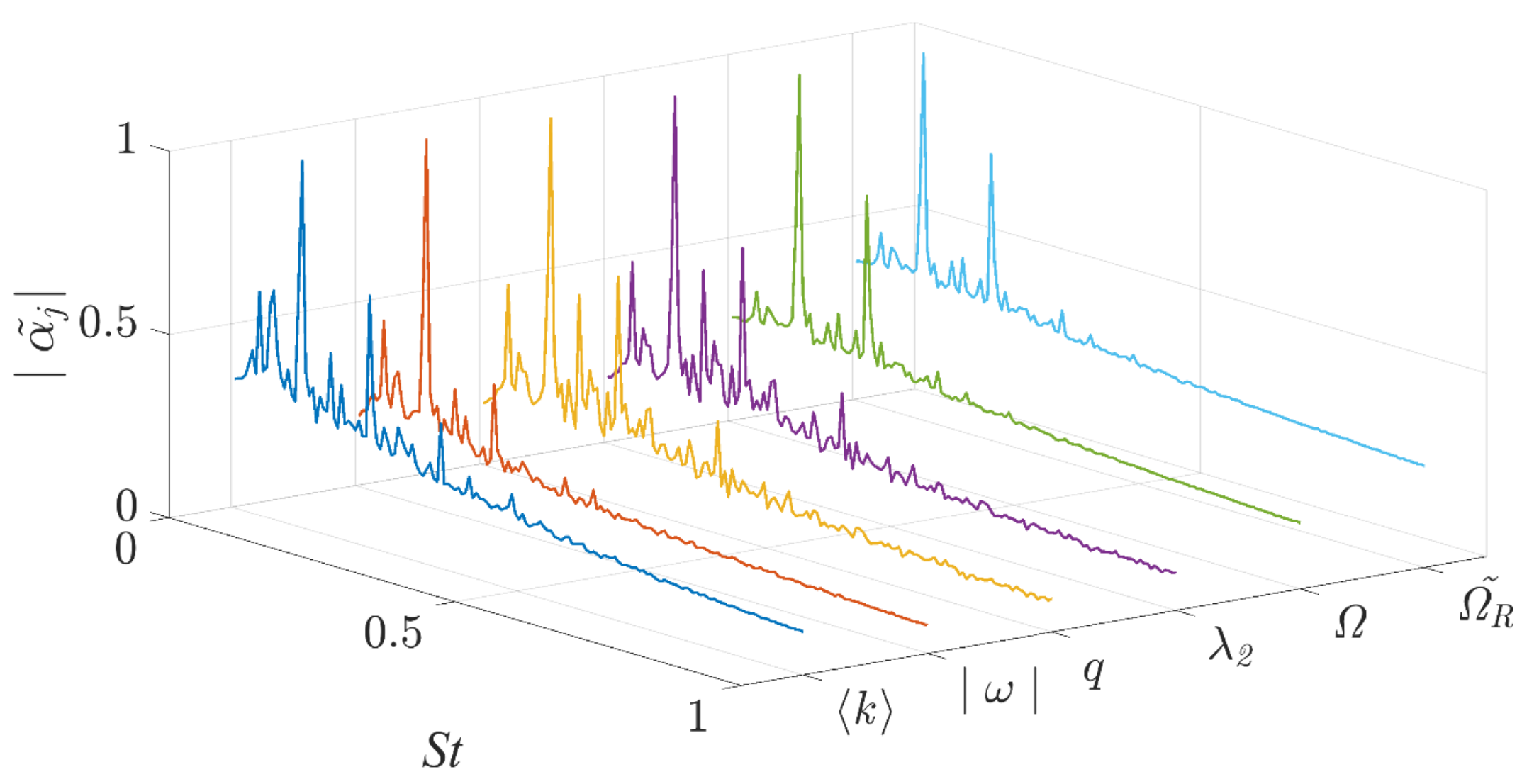

(a)

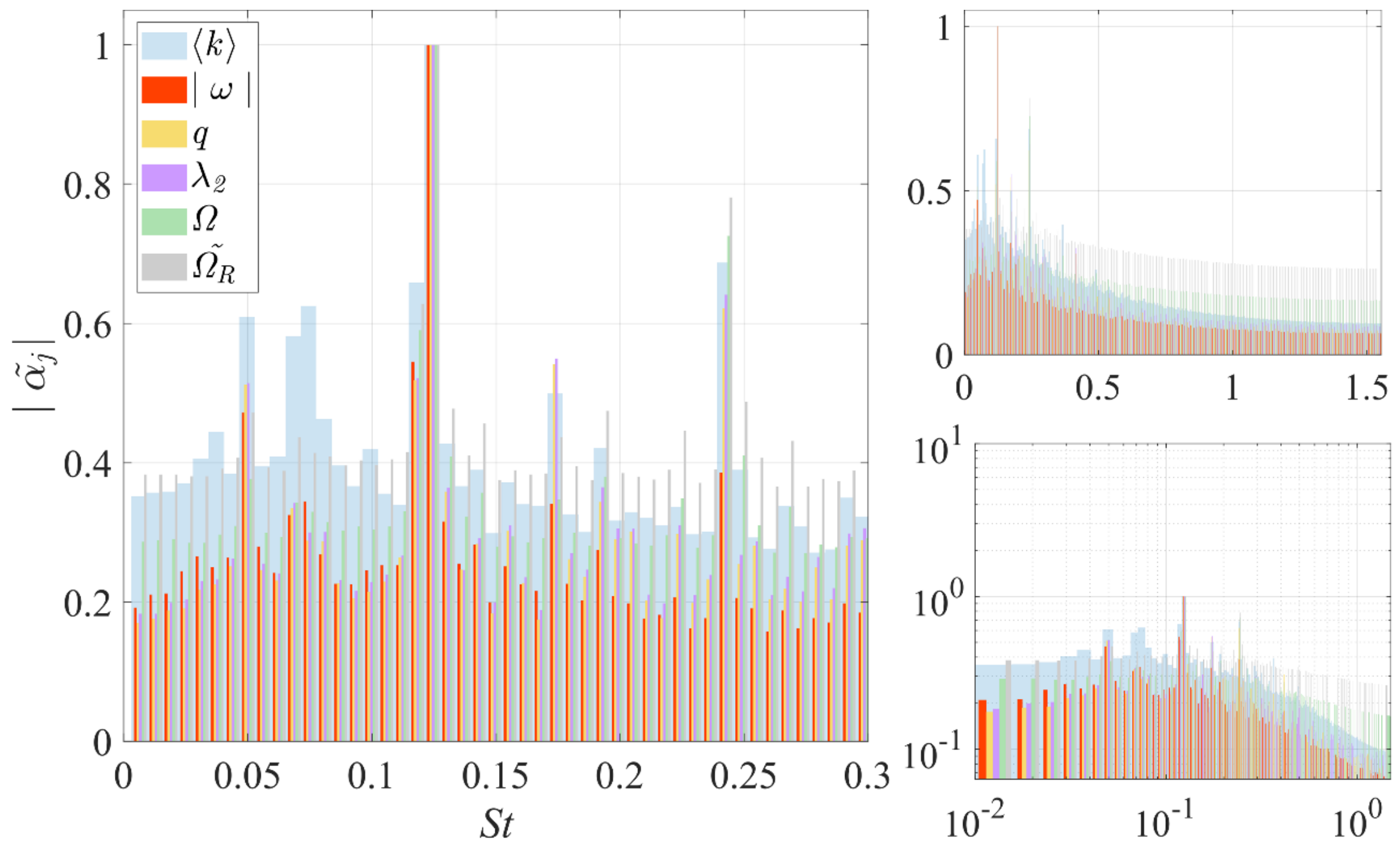

(b)

Figure 12: (a) $|\tilde{\alpha}_j|$ versus *St* of the turbulence kinetic energy and vortex identification criteria ($\langle k \rangle$, $|\omega|$, $q$, $\lambda_2$, $\Omega$, $\tilde{\Omega}_R$); (b) discrete Koopman modes in linear scale and *St=0-0.3* (left), linear scale and *St=0-1.5* (right top), and log scale and *St=0-1.5* (right bottom).

## *4.2 Inter-group constitution*

The intra-group dynamics, while pinpointing the singularities of *w* and the Reynolds stresses, unveiled a remarkable consensus between the natural structures of fluid excitation and structural response. By studying the inter-group dynamics, this section confirms the constitutive relationship to underscore a significant step towards understanding FSI.

### *4.2.1 Fluid-structure correspondence*

To begin, figure 13a presents the $|\tilde{\alpha}_j|$ versus $St_j$ spectra of representative wall and field measurables. To facilitate a clearer message, we also present the ten most dominant modes of all 18 LTI models on the discretized spectra. In figure 13b, colors (blue, orange, green, and maroon) distinguish the four groups, and the color darkness and marker radius figuratively illustrate the spatiotemporal dominance of a mode. Previous conclusions are reaffirmed: except for *w* and the Reynolds stresses, all other measurables reflect the primary peak at *St=0.1242*, the secondary peak at *St=0.0497*, and the ancillary peaks to different degrees.

Most importantly, the direct constitutive relationship between the fluid and structure is lucid by mere inspection. The responses of the on-wind walls are primarily excited by the broadband primary peak at *St=0.1242* and the narrowband secondary peak at *St=0.0497*. The correspondence is highlighted in sky blue and named *Class 1*. The structural response of the downstream wall, while still dominated by the primary peak, reflects several subsidiary excitations, namely the ancillary peaks of descending dominance at *St=0.1739, St=0.0683, St=0.1925, and St=0.2422*. The correspondence is highlighted in lavender and named *Class 2*. Other peaks from figure 13a without unanimous agreements from figure 13b are filtered out.

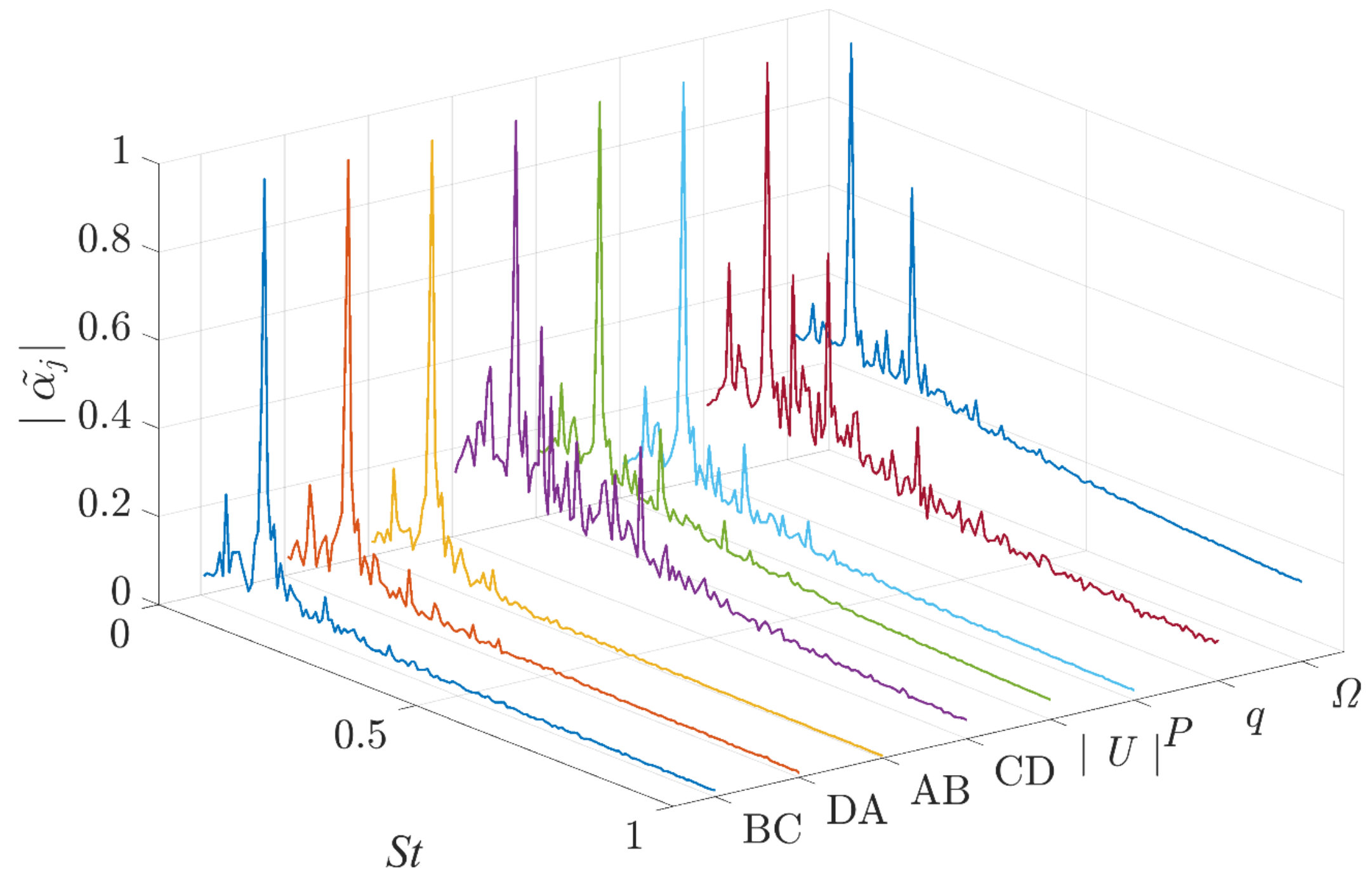

1
0.8
0.6
0.4
0.2
0
0
0.5
1
St
BC
DA
AB
CD
| U |
P
q
Ω

(a)

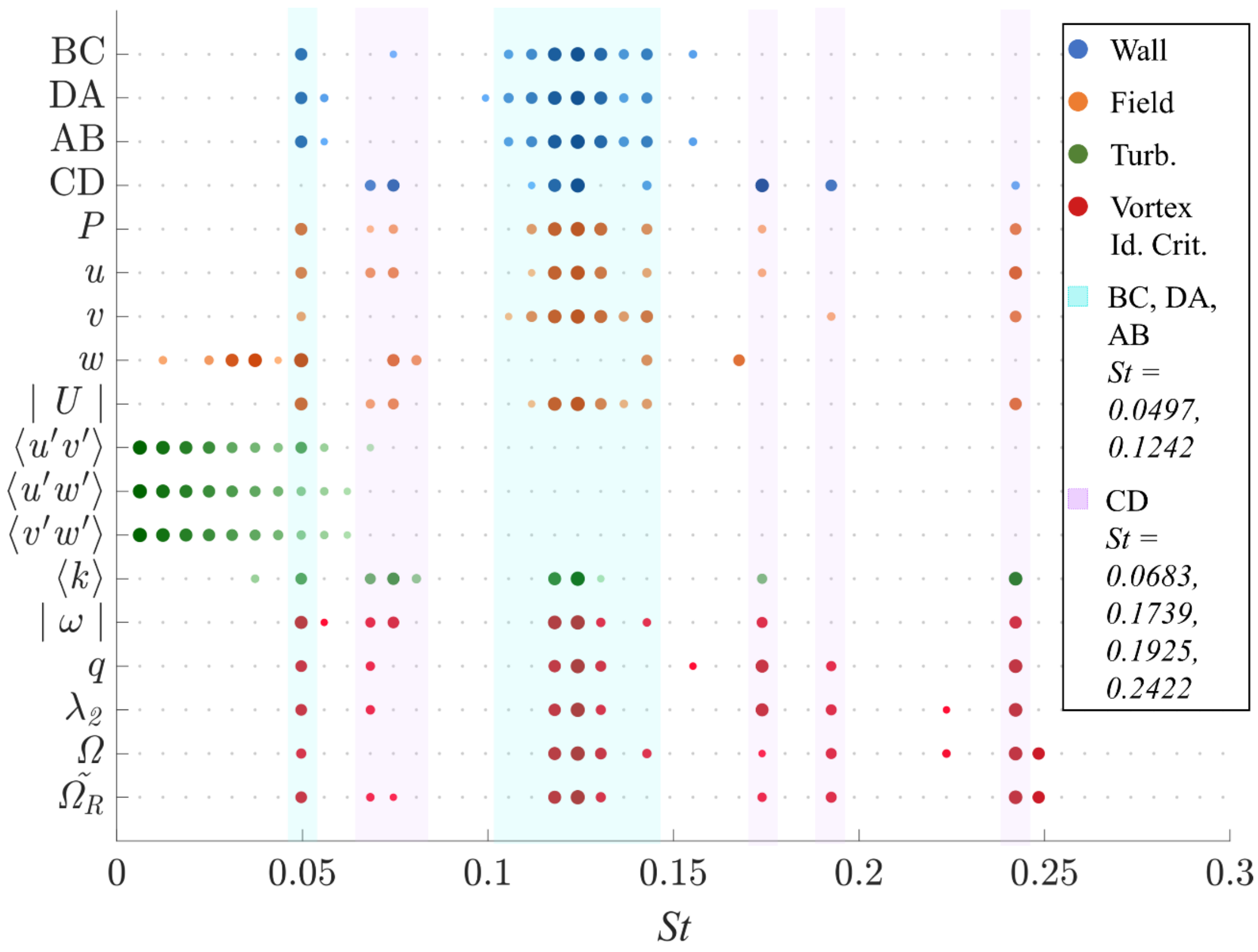

BC
DA
AB
CD
P
u
v
w
| U |
⟨u′v′⟩
⟨u′w′⟩
⟨v′w′⟩
⟨k⟩
| ω |
q
λ2
Ω
Ω̃R
0
0.05
0.1
0.15
0.2
0.25
0.3
St
Wall
Field
Turb.
Vortex Id. Crit.
BC, DA, AB St = 0.0497, 0.1242
CD St = 0.0683, 0.1739, 0.1925, 0.2422

(b)

Figure 13: (a) $|\tilde{\alpha}_j|$ versus *St* of structure (BC, DA, AB, CD) and representative field ($|U|$, *P*, *q, Ω*) measurables; (b) 10 modes with the highest $|\tilde{\alpha}_j|$ versus *St* of the 18 Koopman-LTI systems.

#### *4.2.2 Implications and Significance*

Unearthing from the observations is a crucial revelation. In the moderate subcritical regime during the shear layer transition II, the morphology of the prism wake, allegedly entangled by millions of dynamical components, reduces to only six dominant contributors (*Class 1* and *Class 2*). The three geometrically distinct on-wind walls converge to a single, dynamically unified fluid-structure interface, which is excited by only two predominant mechanisms. As a distinct interface, the downstream wall is excited by a whole different class of four fluid mechanisms. The complete revelation of the prism wakes comes down to the issue of understanding the six mechanisms.

The *a posteriori* evidence also facilitates the notable capacity of the Koopman-LTI. It substantiated our intuition that certain laws and principles relate and govern the fluid and structure, extract them from data, and store them explicitly in the linear, time-invariant mapping matrix. Though solving the governing equations is still a whole different story, the tangible possession is already a huge step forward.

The practical realization of the Koopman-LTI is also straightforward by the procedure from figure 1. The staple is the linear-time-invariance notion. With it, the input data can be independently sampled, the algorithm can be as simple as the vanilla DMD, and identifying the fluid-structure constitution is as easy as inspection.

### *4.2.3 Disambiguation*

On a different note, several other intriguing observations arise from figure 13. Among the vortex fields, $\langle k \rangle$ and $|\omega|$ miss, though to different degrees, some ancillary peaks, such as the ones at *St=0.1739* or *0.1925*. However, the second- and third-generation vortex fields capture all the pertinent modes. Besides making them the optimal identifiers, the observation also suggests that structure responses are closely associated with, if not directly instigated by, vortex dynamics. In practice, if one finds the calculations of vortex fields cumbersome or their interpretations less intuitive, the $|U|$ and $P$ fields generally suffice. The adequacy of $u$ in the present work is attributed to the dominance of convection, therefore, case-specific.

Some may argue that the user-defined threshold of *ten* modes might oversimplify the dynamics, artificially constraining the configuration to only six dominant mechanisms. The doubt is reasonable and deserves clarification. Figure 13a presents the untruncated spectra, and we further supplement figure 14 with 20 and 50 dominant modes. Per an increase from 10 to 20, the most apparent change is the widening of existing peaks, symbolizing the enhanced broadband content (figure 13a). Moreover, some *Class 2* mechanisms begin to show, though minor in extent, effects on the on-wind walls, for example, the ones at *St=0.0683*, *0.1739*, and *0.2422*.

The new additions are highlighted in gold. There are no new, isolated peaks on the on-wind walls. Perhaps only the downstream response at *St=0.2236* qualifies because four vortex fields illustrate its presence. However, its dominance is relatively bleak. The other two peaks at *St=0.2987* and *0.3664* are only sporadically identified by individual measurables. To this end, the six dominant mechanisms are the most prevailing excitations of the prism wake.

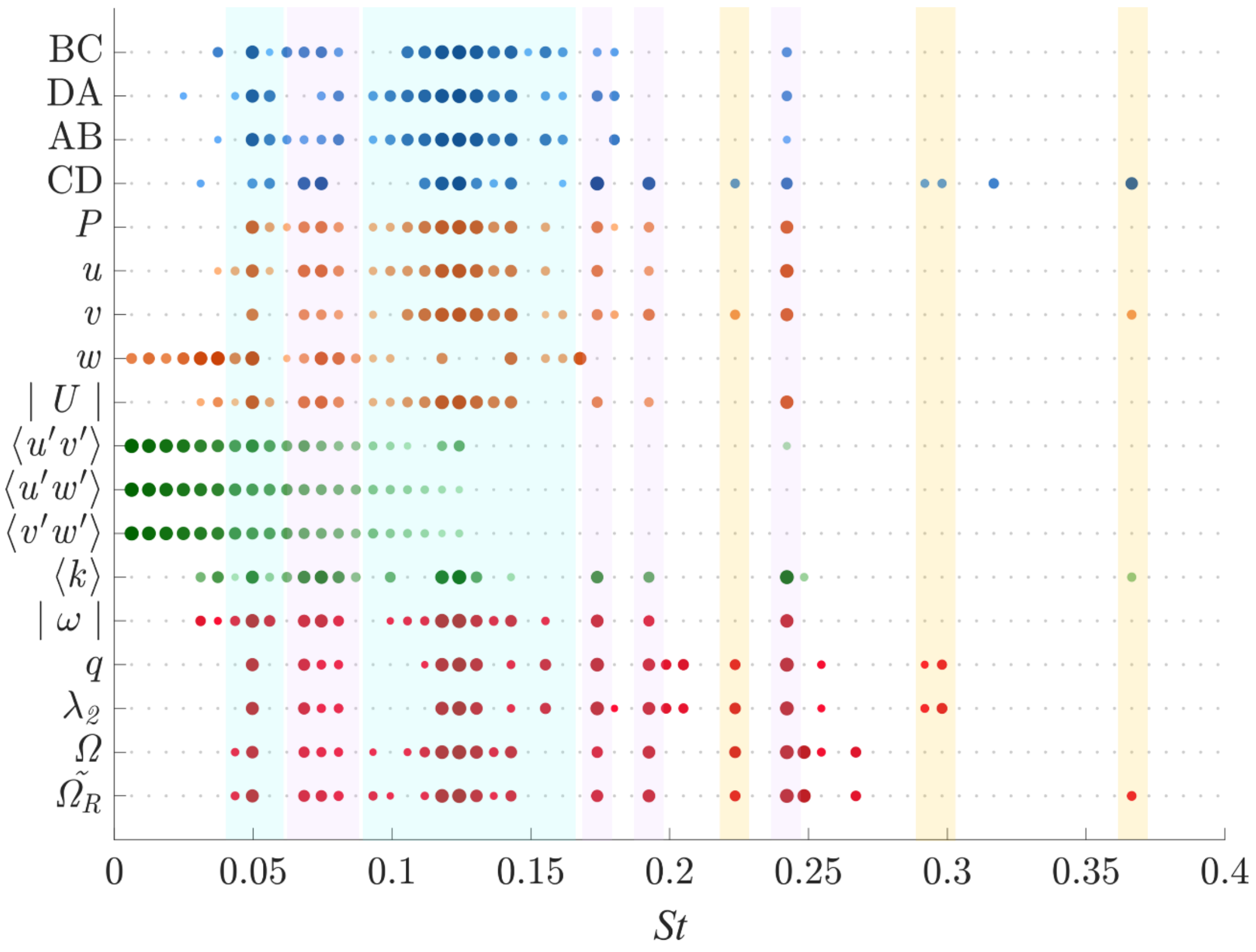

(a)

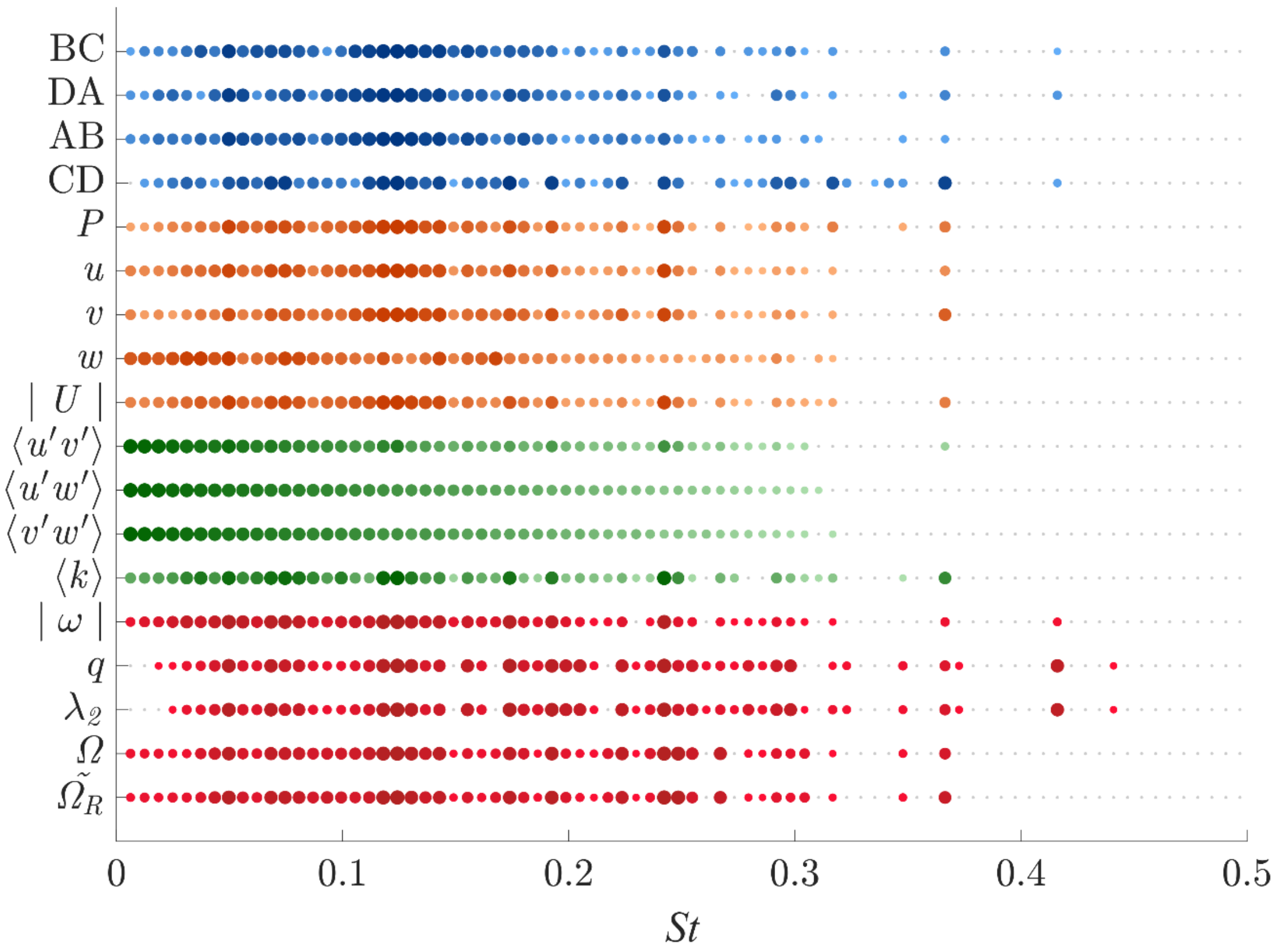

(b)

Figure 14: (a) 20 most dominant modes and (b) 50 most dominant modes versus *St* of the 18 Koopman-LTI systems.

The analysis also exposes another intriguing point. Some mechanisms, like the one at *St=0.3664*, though deemed as trivial by $|U|$ (ranks the 27th), $q$ (ranks the 29th), and $\tilde{\Omega}_R$ (ranks the 15th), projects an overwhelming influence on the downstream wall (ranks the 6th). The opposite is also true. For example, the mechanism at *St=0.2484* is deemed prominent by both $\Omega$ and $\tilde{\Omega}_R$ (ranks the 4th), but instigates only trivial reactions from the structure (ranks >25 for all walls). Even after the global linearization, the interactive mechanisms between fluid and structure are still anything but simple. The transfer of energy demands further investigation.

Thereafter, figure 14b illustrates the change when the dominant modes are increased to 50. All measurables unanimously include the low-frequency spectrum. This is a consensus on the dominance of the low-wavenumber, more energetic eddies. Even for outliers like $w$ and the Reynolds stresses, turbulence's energy preference and natural structures clearly persist. The universal image also attests to the capture of all the important dynamics.

#### *4.2.4 Connection to Fourier Transform*

The preceding analysis illustrate some resemblance between the Koopman and the Fourier analyses. For zero-mean data taken from linearly independent snapshots, Chen et al. (2012) had formally formulated the mathematical relationship between the DMD and the Discrete-Fourier transform (DFT). Although real flows can rarely meet this condition, but as pinpointed by Towne et al. (2018):

> *Mezić* (2005) *showed that for any dynamical system with a Borel probability measure, the growth/decay rate is zero and Koopman modes*

*are equivalent to Fourier modes. Stationary flows possess an ergodic measure by definition, so their Koopman modes are simply Fourier modes.*

The stationary, mean-subtracted data in this work mimics the perennially oscillatory Koopman modes, and the non-zero growth/decay rate is the artifact of the approximation. As seen in figure 8, the numerical zero growth/decay rate ($O^{-12}$ ) suggests that the DMD, Koopman, and DFT modes are practically identical.

The mathematical connection motivated the subsequent empirical comparison between the Koopman-LTI and the power spectral density (PSD) analyses in figure 15. This work performs the standard PSD with seven data series, four pressure series at the geometric center of each wall and another three *u* series at Points 1, 2, and 3, which characterize stagnation, shear layers, and the wake, respectively (see figure 3). In some respects, extracting pointwise series from a full-scale, three-dimensional numerical field is a cumbersome process *per se*.

The typical PSD characterization is local, so the capture of a system-wise representative trend is somewhat fortuitous. Apart from BC and DA, the primary peak at $St = 0.1242$ is unclear for all other five series (see figure 15b). As readers may sympathize, the PSD in practice is often trial-and-error based and depends heavily on data selection, hence a user’s experience. Due to these limitations, one can hardly deduce a fluid-structure constitution from it. The visualization of Fourier modes is also uneasy. The Koopman-LTI bypasses the constraints of locality, esotericism, and visualization by generating a globally optimal characterization. However, the price is an exponential increase in computation cost. On this note, we clarify that the intention is not to adjudicate the superiority of one technique over the other. We merely advise users to make informed decisions about the compromises between locality and computation.

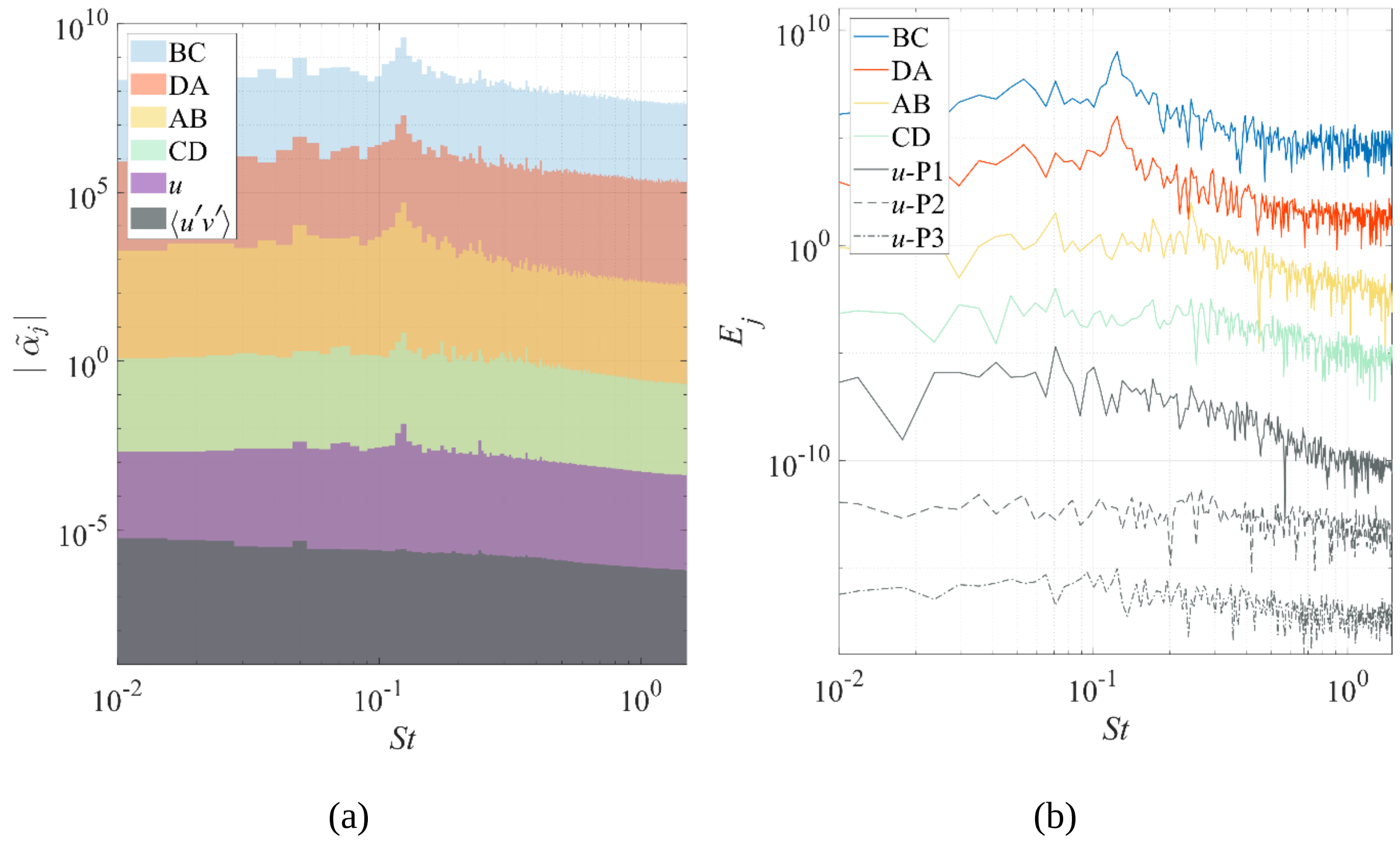

Figure 15: (a) $|\tilde{\alpha}_j|$ versus *St* of selected Koopman-LTI systems on log scale; (b) power-spectral analysis of selected data series of prism walls and *u*-velocity field.

# 5. Conclusions

This work focused on the analytical end of the Koopman analysis. We proposed the Linear-Time-Invariance (LTI) notion, or the Koopman Linearly-Time-Invariant (Koopman-LTI) modular architecture, to study fluid-structure interactions. In the pedagogical demonstration on the prism wake, the LTI models, generated from 18 structure and field measurables, captured all the recurring dynamics, so the Koopman linearization yielded remarkably trivial mean and rms errors of $O^{-12}$ and $O^{-9}$, respectively. With the LTI notion, the vanilla DMD also accurately approximated the Koopman modes, rendering an error of $O^{-8}$. To this end, the LTI models provided a near-exact linearization of the original nonlinear fluid dynamics.

This work also facilitated a pathway to deterministically associate the fluid and structure. By the LTI models, the prism wake undergoing the shear layer transition II was reduced to only six dominant excitation-response mechanisms. The fluid-structure constitution classified the upstream and crosswind walls as a single interface, which is dominated by merely two mechanisms at $St=0.1242$ and $0.0497$. The downstream wall is a distinct interface and is dominated by four other mechanisms at $St=0.0683$, $0.1739$, $0.1925$, and $0.2422$. The solution of the prism wake essentially comes down to the matter of understanding the six mechanisms, which Part 2 will address.

The spectral content of the 18 measurables has been elucidated. Apart from the dynamical independence of the downstream wall, the z-component velocity is also trivial in this configuration. The velocity and pressure fields are convection-dominated. The Reynolds stresses describe the energy of eddies, so they are irrelevant for system characterization. The turbulence kinetic energy matches the vortex fields, showing the close association of vortex dynamics to dilation or their indifference to distortion. The vortex fields also best characterize the wall pressures, attributing structure responses to the origin of vortex activities.

The merit of the Koopman-LTI most avidly reflects on its data-driven nature and modular architecture. It accommodates all data types and Koopman algorithms. Our success with the most rudimentary DMD on inhomogeneous anisotropic turbulence also attests to its broad replicability. Finally, this work highlights the equal importance of the Koopman analysis's analytical end, which is often overshadowed by the zeal toward algorithmic development.

# Acknowledgements

We give a special thanks to the IT Office of the Department of Civil and Environmental Engineering at the Hong Kong University of Science and Technology. Its support for installing, testing, and maintaining our high-performance servers is indispensable for the current project.

## Funding

The work described in this paper was supported by the Research Grants Council of the Hong Kong Special Administrative Region, China (Project No. 16207719), the Fundamental Research Funds for the Central Universities of China (Project No. 2021CDJQY-001), the National Natural Science Foundation of China (Project No. 51908090 and 42175180), the Natural Science Foundation of Chongqing, China (Project No. cstc2019jcyj-msxmX0565 and cstc2020jcyj-msxmX0921), the Key project of Technological Innovation and Application Development in Chongqing (Project No. cstc2019jscx-gksbX0017), and the Innovation Group Project of Southern Marine Science and Engineering Guangdong Laboratory (Project No. 311020001).

## Conflict of Interest

The authors declare that they have no conflict of interest.

## Availability of Data and Material

The datasets generated during and/or analyzed during the current work are restricted by provisions of the funding source but are available from the corresponding author on reasonable request.

## Code Availability

The custom code used during and/or analyzed during the current work are restricted by provisions of the funding source.

## Author Contributions

All authors contributed to the study conception and design. Funding, project management, and supervision were led by Tim K.T. Tse and Zengshun Chen and assisted by Xuelin Zhang. Material preparation, data collection, and formal analysis were led by Cruz Y. Li and Zengshun Chen and assisted by Asiri Umenga Weerasuriya, Yunfei Fu, and Xisheng Lin. The first draft of the manuscript was written by Cruz Y. Li and all authors commented on previous versions of the manuscript. All authors read, contributed, and approved the final manuscript.

## Compliance with Ethical Standards

All procedures performed in this work were in accordance with the ethical standards of the institutional and/or national research committee and with the 1964 Helsinki declaration and its later amendments or comparable ethical standards.

## Consent to Participate

Informed consent was obtained from all individual participants included in the study.

## Consent for Publication

Publication consent was obtained from all individual participants included in the study.